\documentclass{jfm}

\pdfoutput=1

\usepackage{graphicx}
\usepackage{newtxtext}
\usepackage{newtxmath}
\usepackage{multicol}
\usepackage{natbib}
\usepackage{amsmath}
\usepackage{xcolor}
\usepackage{hyperref}
\usepackage[export]{adjustbox}
\hypersetup{
    colorlinks = true,
    urlcolor   = blue,
    linkcolor  = blue,
    citecolor  = blue,
}

\newcommand{\RomanNumeralCaps}[1]

\definecolor{colorC0}{RGB}{31,119,180}
\definecolor{colorC1}{RGB}{255,127,14}
\definecolor{colorC2}{RGB}{44,160,44}
\newcommand{\rev}[1]{{#1}}

\nonstopmode{} 

\shorttitle{Reconstruction of free-surface flows based on CNN}
\shortauthor{A. Xuan and L. Shen}

\title{Reconstruction of three-dimensional turbulent flow structures using surface measurements for free-surface flows based on a convolutional neural network}

\author{Anqing Xuan \and Lian Shen\corresp{\email{shen@umn.edu}}}
\affiliation{%
Department of Mechanical Engineering and Saint Anthony Falls Laboratory, University of Minnesota, Minneapolis, Minnesota 55455, USA
}%

\newcommand{\sublabel}[1]{{\textit{#1}}}

\begin{document}

\maketitle

\begin{abstract}
A model based on a convolutional neural network (CNN) is designed to reconstruct the three-dimensional turbulent flows beneath a free surface using surface measurements, including the surface elevation and surface velocity. Trained on datasets obtained from the direct numerical simulation (DNS) of turbulent open-channel flows with a deformable free surface, the proposed model can accurately reconstruct the near-surface flow field and capture the characteristic large-scale flow structures away from the surface. The reconstruction performance of the model, measured by metrics such as \rev{the normalised mean squared reconstruction errors and scale-specific errors}, is considerably better than that of the traditional linear stochastic estimation (LSE) method. We further analyse the saliency maps of the CNN model and the kernels of the LSE model and obtain insights into how the two models utilise surface features to reconstruct subsurface flows. \rev{The importance of different surface variables is analysed based on the saliency map of the CNN, which reveals knowledge about the surface--subsurface relations.} The CNN is also shown to have a good generalization capability with respect to the Froude number if a model trained for a flow with a high Froude number is applied to predict flows with lower Froude numbers. The results presented in this work indicate that the CNN is effective regarding the detection of subsurface flow structures and \rev{by interpreting the surface--subsurface relations underlying the reconstruction model,} the CNN can be a promising tool for assisting with the physical understanding of free-surface turbulence.
\end{abstract}



\section{\label{sec:intro}Introduction}
The motions of the free surface of a water flow can exhibit various signatures, such as waves, ripples and dimples~\citep{sarpkaya1996,brocchini2001}, which are influenced by the flow underneath and governed by the kinematics and dynamics of the free surface. Inferring information about subsurface flows and even reconstructing the subsurface flow field from observable free-surface features are of great interest for many applications, \rev{such as non-intrusive flow measurements. For example, measurements of the turbulence in the ocean boundary layer can be challenging in the field. In situ measurements relying on single or multiple devices, such as the acoustic Doppler current profiler, can only provide sampled or averaged statistics about the subsurface motions. The capability of inferring the subsurface flow field from free-surface motions can enable measuring subsurface flows through remote sensing and aerial imaging techniques, which have already been applied to the measurements of surface currents~\citep[see e.g.][]{lund2015,metoyer2021} and can provide a greater spatial coverage than point measurements. By identifying subsurface flow structures, which may originate from submerged objects, subsurface flow reconstruction can also benefit applications such as bathymetry mapping and the detection of submerged objects.}

In turbulent free-surface flows, the motion of the free surface is related to the various turbulent coherent structures located underneath. For example, in turbulent open-channel flows, it has been found that the turbulence eddies generated near the water bottom can rise to the near-surface region and excite the free surface~(e.g.\ \citealp*{komori1989}; \citealp{komori1993,rashidi1997}; \citealp{muraro2021}). \citet{nagaosa1999} and~\citet{nagaosa2003} tracked the evolution of hairpin-like vortices from the bottom wall to the near-surface region and provided detailed visualisations of the vortices impacting the free surface and the vortex-induced surface velocities. It was found that upwelling vortices can induce spiral motions at the free surface~\citep*{pan1995,kumar1998}. These studies advanced our understanding of the interactions between free surfaces and turbulence. However, estimating subsurface flows remains challenging despite our knowledge of the correlations between certain surface signatures and flow structures, such as the relation between surface dimples and surface-connected vertical vortices. This is because turbulence is characterised by nonlinear processes and flow structures with various types and scales, which result in complex surface manifestations. Furthermore, water waves, which are governed by the free-surface kinematic and dynamic boundary conditions, are ubiquitous in free-surface flows and can arise from disturbances caused by turbulence. For example, random capillary--gravity waves can be excited by turbulence eddies~\citep{savelsberg2008}. Isotropic small-scale waves were also observed in the numerical experiments conducted by~\citet{yamamoto2011}. These oscillatory motions may further obfuscate the correlations between free-surface features and subsurface structures.

In previous research, some efforts have been made to infer subsurface characteristics from surface features. \citet{koltakov2013} showed that the mean depth and the effective roughness of the bottom can be inferred from surface motions. \citet{mandel2017} proposed an algorithm based on the Schlieren method that can track the scales and movements of Kelvin--Helmholtz rollers over submerged canopies. In these works, the predictable information was limited to either targeted features, such as Kelvin--Helmholtz rollers, or statistical features, such as bottom roughness.

Inferring flow information from boundary observations is also of interest for other types of flows. For example, for wall-bounded flows, various techniques for reconstructing flow fields from wall shear stress and/or pressure have been developed, including the Kalman filtering technique~\citep*{colburn2011}, the linear stochastic estimation (LSE) method~\citep{suzuki2017,encinar2019} and the adjoint method~\citep{wang2021a}. These methods are capable of estimating three-dimensional flow velocities, which contain more details than what can be inferred by the aforementioned techniques for free-surface flows. How turbulence, a chaotic dynamical system, impacts the accuracy of reconstruction algorithms has also been studied in detail for wall-bounded flows. For example, based on an adjoint-variational reconstruction method,~\citet*{wang2022} quantitatively studied the difficulty of flow reconstruction by computing the sensitivity of the measurements to the flow state and showed how the wall distance and measurement time affect the reconstruction accuracy.

In recent years, various machine learning methods based on neural network (NN) architectures have seen increasing use in many fields involving fluid dynamics and turbulence research. NNs are inspired from the biological networks of neurons and use networks of artificial neurons to approximate the nonlinear relationships between input and output data~\citep*{duraisamy2019,brunton2020}. Specialised NNs have also been developed for various tasks, e.g. the convolutional NNs (CNNs)~\citep[see, e.g.][]{lecun1998} are commonly used for processing image-like data, and recurrent NNs (RNNs)~\citep[see, e.g.][]{rumelhart1986} are employed for temporal data. Successful applications of NNs in fluid dynamics include dimensionality reduction (\citealp*{milano2002,murata2020a,fukami2020,page2021}), turbulence modelling~\citep*{ling2016,parish2016,wang2017}, flow control and optimisation~\citep{park2020}, and prediction of the flow evolution~\citep{lee2019a,srinivasan2019a,du2021}. Machine learning has also been applied to the identification of special regions in fluid flows, such as identifying the turbulent/non-turbulent regions~\citep{li2020a} and finding the dynamically significant regions in turbulent flows~\citep*{jimenez2018,jagodinski2020}. In addition, NNs have been found useful in inverse problems. For example, it has been shown that NNs can infer small-scale structures and reconstruct turbulent flow fields from low-resolution spatio--temporal measurements (\citealp{xie2018,werhahn2019}; \citealp*{fukami2019a}; \citealp{liu2020a}; \citealp*{fukami2021}; \citealp{kim2021}). This type of super-resolution reconstruction technique can be used to improve the quality of experimental and simulation data with limited resolutions.

NNs have also been successfully applied to the state estimation of turbulent flows from indirect or limited measurements. \citet{milano2002} proposed a fully connected NN model that reconstructs the near-wall velocity field of a turbulent channel flow using wall shear stress and pressure. They showed that the NN with nonlinear mappings produced better reconstructions than the linear model. \citet*{guemes2019} proposed a CNN model combined with proper orthogonal decomposition to reconstruct large-scale motions from wall shear stress.  A generative adversarial network was developed by~\citet{guemes2021} to reconstruct a flow from wall shear stress and pressure. \citet{guastoni2020} showed that a fully-convolutional NN could also produce good velocity field reconstructions from wall shear stress. This model was further extended by~\citet{guastoni2021} to predict three-dimensional velocity fields. \citet{erichson2020} developed a shallow NN to reconstruct a flow field from a limited number of sensors and showed that it performed well for a variety of problems, including the flow past a cylinder, the distribution of sea surface temperatures, and the homogeneous isotropic turbulence. \citet{matsuo2021} combined two-dimensional and three-dimensional CNNs to reconstruct a three-dimensional flow field from several two-dimensional slices. The known physical laws can also be integrated with NNs for reconstruction purposes. For example, the governing equations that describe the evolution of flows were used by~\citet*{raissi2019} to build a physics-informed NN, which can reconstruct velocity and pressure fields using only the visualisation of the mass concentration in the corresponding flow~\citep*{raissi2020}. 

In the present study, we aim to explore the feasibility of using a CNN to reconstruct a turbulent flow under a free surface based solely on surface observations. \citet*{gakhar2020} developed a CNN to classify bottom roughness types from the surface elevation information, indicating that machine learning can be applied to inverse problems involving free-surface flows. In our work, the reconstruction method directly predicts three-dimensional velocity fields, which is a significant improvement over the aforementioned subsurface flow inference techniques that can only predict targeted features. Although similar three-dimensional reconstruction models have been proposed for turbulent channel flows, the available reconstruction methods for free-surface flows are limited. Moreover, compared to reconstructions from wall measurements, reconstructions from free surfaces have unique challenges associated with surface deformation and motions.
In addition to demonstrating this CNN application, the present work also discusses the implications on the flow physics underlying the interactions between a free surface and turbulence. 

The CNN model developed in the present work takes the surface elevation and surface velocities as inputs to estimate the three-dimensional velocity field underneath a free surface.  The model embeds the fluid incompressibility constraint and is trained to minimize the induced reconstruction errors using the data obtained from the direct numerical simulation (DNS) of turbulent open-channel flows. For comparison, a reconstruction model based on the LSE method is also considered. Both qualitative and quantitative measures indicate that near the surface, both the CNN and LSE models can reconstruct the flow fields reasonably well; however, away from the surface, the CNN model achieves significantly better reconstruction performance than the LSE model. We further evaluate the models' reconstruction performance at different spatial scales to provide a comprehensive picture of what flow structures can be predicted from a free surface and to what extent.

The present study also aims to gain insights into how the LSE and CNN models are related to the flow physics of subsurface turbulence. For the LSE model, its linear transformation kernel is examined, revealing characteristic coherent structures that are linearly correlated with the surface motions. For the CNN model, we compute saliency maps to assess the importance of each surface quantity to the reconstruction outcomes and find that some quantities are less important and can be omitted with only a negligible impact on the reconstruction accuracy. The relative importance levels of different surface variables are found to depend on the Froude number as a result of the influence of gravity on the free-surface dynamics. We also find that the salient regions of the CNN have correlations with dynamically important structures beneath the surface, which indicates that the CNN can identify their manifestations on the free surface and utilise them to reconstruct the flow field. We consider this outcome a first step towards interpreting the outstanding performance of the CNN model.
We also apply the CNN model that is trained using data for one Froude number to flows with other Froude numbers to assess the generalization ability of the CNN model with respect to free-surface flows with various surface deformation magnitudes.

The remainder of this paper is organized as follows. In~\S\,\ref{sec:method}, we present the formulation of the free-surface flow reconstruction problem and the proposed methods. In~\S\,\ref{sec:result}, we assess the performance of the reconstruction models. In~\S\,\ref{sec:discussions}, we further discuss the obtained physical insights regarding these reconstruction models and their relations with flow dynamics. In~\S\,\ref{sec:conclusions}, we summarise the results of the paper.

\section{\label{sec:method}Methodology}
The reconstruction of a subsurface flow field based on surface observations is equivalent to finding a mapping $\mathcal{F}$ from the surface quantities to a three-dimensional velocity field:
\begin{equation}
    \tilde{\boldsymbol{u}} = \mathcal{F}(\boldsymbol{E}), \label{eq:mapping}
\end{equation}
where $\boldsymbol{E}$ denotes a vector consisting of the considered surface quantities, and $\tilde{\boldsymbol{u}}=(\tilde{u}, \tilde{v}, \tilde{w})$ denotes the estimated velocity, with $\tilde{u}$, $\tilde{v}$ and $\tilde{w}$ being the components of $\tilde{\boldsymbol{u}}$ in the $x$-, $y$- and $z$-directions, respectively. We denote the horizontal directions as the $x$- and $y$-directions and the vertical direction as the $z$-direction. In the present study, the elevation and velocity fluctuation observations at the surface are used as the inputs of the reconstruction process, i.e.
\begin{equation}
    \boldsymbol{E} = (u_s, v_s, w_s, \eta), \label{eq:surfaceE}
\end{equation}
where the subscript ${(\cdot)}_s$ denotes the quantities evaluated at the free surface and $\eta(x,y,t)$ denotes the surface elevation. For reconstruction purposes, $\mathcal{F}$ should minimize the difference between the predicted velocity field $\tilde{\boldsymbol{u}}$ and the true velocity $\boldsymbol{u}$. We note that in the present study, we only consider the spatial mapping between $\tilde{\boldsymbol{u}}$ and $\boldsymbol{E}$; i.e.\ $\tilde{\boldsymbol{u}}$ and $\boldsymbol{E}$ are acquired at the same time instant.

\subsection{\label{sec:cnn}CNN method}
\begin{figure}
    (\sublabel{a})\\
    \begin{center}
    \includegraphics[width=0.9\textwidth]{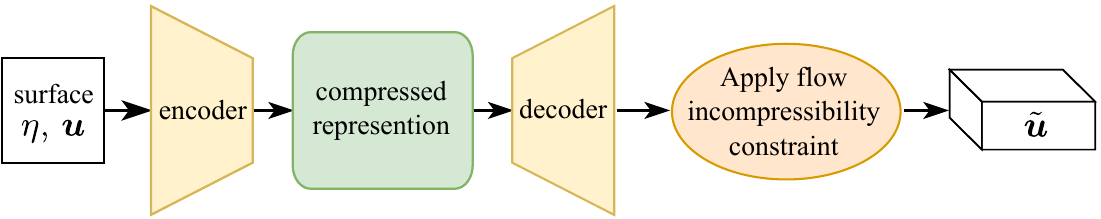}
    \end{center}
    (\sublabel{b})\\
    \begin{center}
        \includegraphics{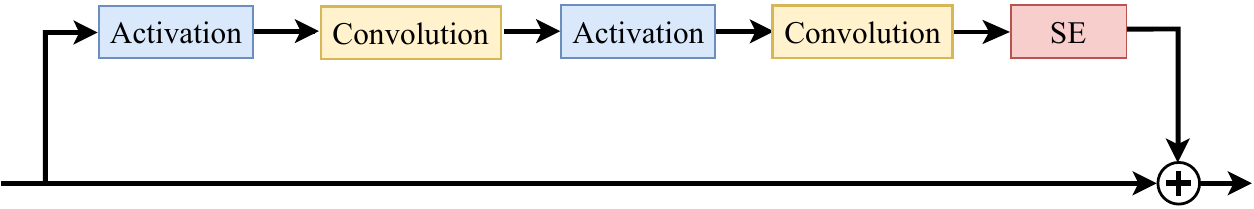}
    \end{center}
    \caption{\label{fig:nn_cells}Building blocks of the network for flow reconstruction. (a) Overview of the reconstruction process. (b) Structure of the residual block in the encoder. SE refers to the squeeze--and--excitation layer~\citep{hu2018}.}
\end{figure}

We utilise a CNN~\citep{lecun1998}, which is widely used to process grid-like data for obtaining their spatial features, to model the mapping $\mathcal{F}$ described above for flow reconstruction purposes. The overall architecture of the NN is sketched in figure~\ref{fig:nn_cells}(\sublabel{a}). The model consists of two parts: an encoder that maps the surface quantities into a representation with reduced dimensions and a decoder that reconstructs the three-dimensional flow field from the reduced-dimension representation. The input fed into the encoder includes the surface quantities $\boldsymbol{E}$ sampled on a grid of size $256 \times 128$. With $\boldsymbol{E}$ consisting of four variables, the total dimensions of the input values are $256 \times 128 \times 4$. The encoding process, which outputs a compressed representation of dimensionality $32 \times 16\times 24$, extracts the most important surface features for the subsequent reconstructions, \rev{which may alleviate the problem of the model overfitting insignificant or nongeneralizable features~\citep{verleysen2005,ying2019}}.
The decoder reconstructs velocities with three components from the output of the encoder onto a three-dimensional grid of size $128\times 64 \times 96$. The total dimensions of the output data are $128\times 64 \times 96 \times 3$. The input and output of the CNN are based on the same time instant, and each instant is considered independently; therefore, no temporal information is used for the reconstruction process.

The detailed architectures of the CNN layers are listed in table~\ref{tab:nn_param}. The basic building block of a CNN is the convolutional layer, which is defined as
\begin{equation}
    X^{out}_n = b_n + \sum_{k=1}^{K^{in}} W_n(k)*X^{in}_k,\; n=1,\dots,K^{out}, \label{eq:conv_layer}
\end{equation}
where $X^{in}_k$ denotes the $k$-th channel of an input consisting of $K^{in}$ channels, $X^{out}_n$ denotes the $n$-th channel of the $K^{out}$-channel output,  $b$ denotes the bias, $W$ is the convolution kernel, and `$*$' denotes the convolution operator. Strictly speaking, `$*$' calculates the cross correlation between the kernel and the input, but in the context of a CNN, we simply refer to it as convolution. We note that in~\eqref{eq:conv_layer}, the input may consist of multiple channels. For the first layer that takes the surface quantities as inputs, each channel corresponds to one surface variable. Each convolutional layer can apply multiple kernels to the input to extract different features and as a result, its output may also contain multiple channels. Each channel of the output is thus called a feature map.

These feature maps are often fed into a nonlinear activation function, which enables the CNN to learn complex and nonlinear relationships. We have compared the performance of several types of nonlinear activation functions, including the rectified linear unit (ReLU) function, the sigmoid function and the hyperbolic tangent function. The hyperbolic tangent function, which performs better, is selected in the present work. We note that~\citet{murata2020a} also found the hyperbolic tangent function performs well for the mode decomposition of flow fields. 

\begin{table}
    \begin{center}
    \begin{tabular}{cccc}
        Input size  &  Operator  & Kernel size & Stride \\
        \hline
        \multicolumn{4}{c}{Encoder} \\
        $256\times 128 \times 4$ &  Convolution  &    $(3,3)$  &  $(1,1)$ \\
        $256\times 128 \times 4$ &  Blur pooling  &    $(3,3)$  &  $(2,2)$ \\
        $128\times 64 \times 16$ & Residual block      &   -  &  - \\
        $128\times 64 \times 16$ &  Convolution     &   $(3,3)$  &  $(1,1)$ \\
        $128\times 64 \times 16$ &  Blur pooling &    $(3,3)$  &  $(2,2)$ \\
        $64\times 32 \times 24$  & Residual block     &   -  &  - \\
        $64\times 32 \times 24$  &  Convolution     &   $(3,3)$  &  $(2,2)$ \\
        $32\times 16 \times 24$  & Residual block     &   -  &  - \\
        \hline
        \multicolumn{4}{c}{Decoder} \\
        $32\times 16 \times 1 \times 24$   & Transposed convolution   &   $(4,4,1)$  &  $(2,2,1)$  \\
        $64 \times 32 \times 1 \times 48$  & Transposed convolution   &   $(3,3,3)$  &  $(1,1,3)$ \\
        $64 \times 32 \times 3 \times 56$  & Transposed convolution   &   $(4,4,4)$  &  $(1,1,2)$ \\
        $64 \times 32 \times 6 \times 32$  & Transposed convolution   &   $(4,4,4)$  &  $(1,1,2)$ \\
        $64 \times 32 \times 12 \times 18$ & Transposed convolution   &   $(4,4,4)$  &  $(1,2,1)$ \\
        $64 \times 64 \times 12 \times 16$ & Transposed convolution   &   $(4,4,4)$  &  $(1,1,2)$ \\
        $64 \times 64 \times 24 \times 16$ & Transposed convolution   &   $(4,4,4)$  &  $(2,1,2)$ \\
        $128 \times 64 \times 48 \times 16$ & Transposed convolution  &   $(4,4,4)$  &  $(1,1,2)$ \\
        $128 \times 64 \times 97 \times 9$ & Convolution   &    $(5,5,5)$  &  $(1,1,1)$ \\
        $128 \times 64 \times 97 \times 3$ & Convolution   &    $(5,5,5)$  &  $(1,1,1)$ \\
        $128 \times 64 \times 97 \times 3$ & Convolution   &    $(4,4,4)$  &  $(1,1,1)$ \\
        $128 \times 64 \times 97 \times 3$ & Convolution   &    $(3,3,3)$  &  $(1,1,1)$ \\
        $128 \times 64 \times 97 \times 3$ & Convolution   &    $(1,1,2)$  &  $(1,1,1)$
    \end{tabular}
    \caption{\label{tab:nn_param}Detailed architecture of the CNN with its parameters, including the input size, kernel size, and stride of each block (if applicable). The input size is written as $N_x\times N_y \times N_c$ for two-dimensional data for the encoder or $N_x \times N_y \times N_z \times N_c$ for three-dimensional data for the decoder, where $N_x$, $N_y$ and $N_z$ denote the grid dimensions in the $x$-, $y$- and $z$-directions, respectively, and $N_c$ denotes the number of channels. Note that the output of each block is the input of the next block. The output size of the last block is $128\times 64 \times 96 \times 3$.}
    \end{center}
\end{table}

\rev{In the encoder, we use strided convolution layers and blur pooling layers to reduce the dimensionality of feature maps. In a strided convolution, the kernel window slides over the input by more than one pixel (grid point) at a time, and as a result of skipping pixels, the input is downsampled. A blur pooling layer can be considered a special strided convolution operation with fixed weights. The blur kernel we use here is the outer product of $[0.25,0.5,0.25]$~\citep{zhang2019}, which applies low-pass filtering and downsampling to the input at the same time. The blur pooling can improve the shift-invariance of the CNN~\citep{zhang2019}, which is discussed further below.}
In the encoder part, we also utilise a design called residual blocks, which is empirically known to ease the training processes of NNs~\citep{he2016b}. The structure of a residual block is illustrated in figure~\ref{fig:nn_cells}(\sublabel{b}). The output of the block is the sum of the original input and the output of several processing layers. Following~\citet{vahdat2020}, the residual block in this work consists of two rounds of activation--convolution layers, followed by a squeeze--and--excitation layer that enables the network to learn crucial features more effectively~\citep*{hu2018}. In our experiments, we find that the use of residual blocks in the encoder part can reduce the number of epochs needed to train the network. The residual blocks can also be implemented for the decoder as in~\citet{vahdat2020},but we do not observe any significant differences in the training or the performance of the network.

The reconstruction process of the decoder uses a transposed convolution operation, which is essentially the adjoint operation of convolution. To interpret the relationships between a convolution and a transposed convolution, we first note that a convolution operation can be computed as a matrix-vector product, with the matrix defined by the kernel weights and the vector being the flattened input. By transposing the matrix, one obtains the adjoint operation, i.e. the transposed convolution, associated with the initial convolution operation. It should also be noted that the dimensions of the input and output of a transposed convolution are the reverse of those of the corresponding convolution. In the decoder, strided transposed convolutions has the opposite effect relative to that of the strided convolutions in the encoder, i.e. they are used to upsample the feature maps.

\rev{To ensure that the CNN can better represent physical correlations between the surface and subsurface flows, the CNN architecture is designed to maintain good translation invariance, i.e. the ability to produce equivalent reconstructions when the surface input is shifted in space. Although the convolution operations are shift-invariant by themselves, the downsampling by strided convolutions can produce different results when inputs are shifted~\citep{zhang2019,azulay2019}. From the perspective of the Fourier analysis, downsampling may introduce aliasing errors at high wavenumber modes of the feature maps, which may be amplified by subsequent layers, and as a result, break the translation invariance of the network. Therefore, following~\citet{zhang2019}, we employ the blur pooling layers, which perform antialiasing while downsampling to reduce aliasing errors. The blur pooling layers act as the first two downsampling layers in the encoder (see table~\ref{tab:nn_param}). As reported in appendix~\ref{sec:appendix_translation}, the present network can produce more consistent reconstructions for shifted inputs than a network that only uses strided convolutions for downsampling. The blur pooling is not applied to the last downsampling operation in the encoder because we find that it barely improves the translation invariance but decreases the reconstruction accuracy as the blurring can result in loss of information~\citep{zhang2019}. We also apply periodic paddings in the horizontal directions to reduce the translation-induced errors owing to the edge effect. Furthermore, we apply random translations in the horizontal directions to the training data, known as data augmentation, which is detailed in~\S\,\ref{sec:preprocess}, to help improve the translation invariance~\citep{kauderer-abrams2017}. We note that complete translation invariance is difficult to achieve because nonlinear activation functions can also introduce aliasing effects~\citep{azulay2019}. Nevertheless, our tests indicate that the reconstructions are consistent for shifted inputs and therefore the translation-induced errors should be small.}

\rev{The choice of model parameters, such as the kernel sizes and strides, is often a balance between performance and computational cost. Most existing CNNs use small kernels, with sizes between $3$ and $7$~\citep[see e.g.][]{simonyan2014,he2016b,tan2019,lee2019a}, for computational efficiency. In the present work, we use kernels of sizes varying from $3$ and $5$. For transposed convolutions, the kernel size is chosen to be divisible by the stride to reduce the checkerboard artifacts~\citep{odena2016}.}
\rev{For both strided convolutions and blur pooling layers, we use a stride of $2$, i.e. a downsampling ratio of 2, which is a common choice when processing images~\citep{tan2019} and fluid fields~\citep{lee2019a,park2020}.}
\rev{Similarly, in the decoder, the spatial dimensions are in general expanded by a ratio of $2$, except for one layer where the features maps are expanded by a ratio of $3$ in the vertical direction owing to the fact that the final dimension $96$ has a factor of $3$. Some transposed convolution layers only expand one spatial dimension to accommodate the different expansion ratios needed in different dimensions to obtain the output grid. Although there are many combinations for how the dimension expansions are ordered in the decoder, we find that the network performance is not particularly sensitive to the ordering.}

\rev{The performance of the CNN is also affected by the number of channels (feature maps) in each layer. A network with more layers can in theory provide a greater capacity to describe more types of features. As presented in \S\,1 of the supplementary material, a network with $50\%$ more channels has almost the same performance as the one presented in table~\ref{tab:nn_param}, indicating that the present network has enough number of feature maps for expressing the surface--subsurface mapping. A special layer worth more consideration is the output of the encoder or the input of the decoder, which is the bottleneck of the entire network and encodes the latent features that are useful for reconstructions. The effect of the dimension of this bottleneck layer on the reconstruction performance is studied by varying the number of channels in this layer. It is found that the present size with $24$ layers can retain most surface features necessary for the network to achieve an optimal performance. More detailed comparisons are provided in \S\,1 of the supplementary material.}
\rev{We have also considered some variants of the network architecture to investigate whether a deeper network can improve the reconstruction performance. These modified networks show no significant performance differences from the current model, indicating that the present network already has enough expressivity to describe the mapping for subsurface flow reconstructions. Detailed comparisons are reported in \S\,2 of the supplementary material.}

The entire network is trained to minimize the following loss function:
\begin{equation}
    J = \frac{1}{3 L_z}\sum_{i=1}^3 {\int_z \frac{\iint_{x,y} {\left| \tilde{u}_i - u_i \right|}^2 \,\mathrm{d}x\mathrm{d}y}{\iint_{x,y} u_i^2 \,\mathrm{d}x\mathrm{d}y} \mathrm{d}z}. \label{eq:loss}
\end{equation}
which first computes the mean squared errors normalised by the plane-averaged Reynolds normal stresses for each component and for each horizontal planes, and then averages these relative mean squared errors. This loss function is defined to consider the inhomogeneity of the open-channel flow in the vertical direction and the anisotropy among the three velocity components. Compared to the mean squared error of the velocity fluctuations (equivalent to the error of the total turbulent kinetic energy), the above loss function gives higher weights to the less energetic directions and less energetic flow regions.


The incompressibility constraint is embedded into the network to produce a reconstruction that satisfies mass conservation. \rev{At the end of the decoder, the reconstructed velocity is determined by
\begin{equation}
    \tilde{\boldsymbol{u}} = \bnabla \times \tilde{\boldsymbol{A}}, \label{eq:curl}
\end{equation}
where $\tilde{\boldsymbol{A}}$ is a vector potential estimated by the network. The solenoidality of $\tilde{\boldsymbol{u}}$ is automatically satisfied following the vector identity $\bnabla \cdot (\bnabla\times \tilde{\boldsymbol{A}}) = 0$. It should be noted that the vector potential of a velocity field is not unique, which can be seen from $\bnabla\times\tilde{\boldsymbol{A}} = \bnabla\times(\tilde{\boldsymbol{A}} + \nabla\psi)$ where $\psi$ is an arbitrary scalar function. Considering that $\tilde{\boldsymbol{A}}$ is just an intermediate quantity and it is the velocity field we are interested in, we do not add any constraints on $\boldsymbol{A}$ to fix the choice of $\psi$. In other words, we let the network architecture and the optimization process to freely estimate the vector potential. We find that predicting the velocity field through its vector potential and predicting the velocity directly have no discernable differences in the reconstruction accuracy (see detailed results in \S\,4 of the supplementary material), indicating that the imposed incompressibility constraint does not negatively impact the CNN's reconstruction capability.}

\rev{The spatial derivatives in~\eqref{eq:curl} are computed using the second-order central difference scheme, which can be easily expressed as a series of convolution operations with fixed kernel coefficients. We have also tested the Fourier spectral method to calculate the derivatives in the $x$- and $y$-directions, which barely affect the reconstruction accuracy but significantly increases the training time. This indicates that a higher-order scheme does not provide additional reconstruction accuracy, which we believe is because most reconstructable structures are low-wavenumber structures and the second-order central scheme is adequately accurate for resolving these structures.}

The total number of trainable parameters in the designed network is $279,551$. The network is trained using the adaptive moment estimation with decoupled weight decay (AdamW) optimiser~\citep{loshchilov2019} with early stopping to avoid overfitting. The training process is performed on an NVIDIA V100 GPU\@.

\subsection{\label{sec:lse}LSE method}
For comparisons with the CNN model, we also consider a classic approach based on the LSE method~\citep{adrian1988}, which is commonly used to extract turbulent coherent structures from turbulent flows~\citep*{christensen2001,xuan2019a}. It has also been applied to reconstruct the velocity on an off-wall plane in a turbulent channel flow from wall measurements~\citep{suzuki2017,encinar2019,guastoni2020}. \citet{wang2021} demonstrated that this method can be used to transform a three-dimensional vorticity field into a velocity field.

The LSE method, as the name suggests, estimates a linear mapping from the observations of some known random variables to unknown random variables. In the present problem, the known random variables are the surface quantities $\boldsymbol{E}$ and the unknowns are the estimated $\tilde{\boldsymbol{u}}$. The reconstruction of $\tilde{\boldsymbol{u}}$ by a linear relationship can be expressed as
\begin{equation}
    \tilde{u}_i(\boldsymbol{x}) = \mathcal{L}_{ij}(\boldsymbol{x}; \boldsymbol{x}_s) E_j(\boldsymbol{x}_s),\; i=1,2,3;\, j=1,2,3,4. \label{eq:lse_def}
\end{equation}
where $\mathcal{L}_{ij}$ (written as $\mathcal{L}$ below for conciseness) are linear operators. In other words, \eqref{eq:lse_def} treats the mapping $\mathcal{F}$ in~\eqref{eq:mapping} as a linear function of $\boldsymbol{E}$. Note that two distinct coordinates are contained in the above equation: $\boldsymbol{x}$ and $\boldsymbol{x}_s$. The coordinate $\boldsymbol{x}$ denotes the position to be reconstructed and $\mathcal{L}$ is a function of $\boldsymbol{x}$ because the mapping between $\boldsymbol{E}$ and $\tilde{\boldsymbol{u}}$ should depend on the location to be predicted. The coordinate $\boldsymbol{x}_s$ denotes the horizontal coordinates, i.e.\ $\boldsymbol{x}_s=(x,y)$, which signifies that the observed surface quantity $\boldsymbol{E}(\boldsymbol{x}_s)$ is a planar field. Here, we denote the surface plane as $S$, i.e.\ $\boldsymbol{x}_s \in S$. For each $\boldsymbol{x}$, $\mathcal{L}$ is a field operator on $S$ that transforms $\boldsymbol{E}(\boldsymbol{x}_s)$ into the velocity $\boldsymbol{\tilde{u}}(x)$. The LSE method states that $\mathcal{L}$ can be determined by
\begin{equation}
    \left\langle{E_m(\boldsymbol{r}') E_j(\boldsymbol{x}_s)}\right\rangle \mathcal{L}_{ij}(\boldsymbol{x};\boldsymbol{x}_s) = \left\langle{u_i(\boldsymbol{x}) E_m(\boldsymbol{r}')}\right\rangle,\;  \forall r'\in S;\,m=1,2,3,4 \label{eq:lse_optimize}
\end{equation}
where $\left\langle\cdot\right\rangle$ is defined as averaging over all the ensembles (instants) in the dataset. The above equation is obtained by minimizing the mean squared errors between $\tilde{u}_i(\boldsymbol{x})$ and ${u}_i(\boldsymbol{x})$ over all the ensembles (instants) in the dataset; i.e. the operator $\mathcal{L}$ defined above is the optimal estimation of the mapping $\mathcal{F}$ in the least squares sense. Note that on the left-hand side of~\eqref{eq:lse_optimize}, $\left\langle{E_m(\boldsymbol{r}') E_j(\boldsymbol{x}_s)}\right\rangle$ should be considered a planar field function of $\boldsymbol{x}_s$ when multiplied by the operator $\mathcal{L}$. By varying $\boldsymbol{r}'$ and $m$, one can obtain a linear system to solve for $\mathcal{L}$. Similar to the CNN model in~\S\,\ref{sec:cnn}, the LSE model only describes the spatial correlations between the subsurface and surface quantities.

As pointed out by~\citet{wang2021}, the operator $\mathcal{L}$ in~\eqref{eq:lse_def} can be alternatively written in an integral form as
\begin{equation}
    \tilde{{u}}_i(\boldsymbol{x}) = \iint_S {l_{ij}(\boldsymbol{x}; \boldsymbol{x}_s) E_j(\boldsymbol{x}_s)}\, \mathrm{d}\boldsymbol{x}_s. \label{eq:lse_integral_def}
\end{equation}
Utilising the horizontal homogeneity of the open-channel turbulent flow, it can be easily shown that $l_{ij}$ is a function that is dependent on $\boldsymbol{x}-\boldsymbol{x}_s$ and, therefore, \eqref{eq:lse_integral_def} can be written as
\begin{equation}
    \tilde{{u}}_i(\boldsymbol{x}) = \iint_S {l_{ij}(\boldsymbol{x}-\boldsymbol{x}_s) E_j(\boldsymbol{x}_s)}\, \mathrm{d}\boldsymbol{x}_s. \label{eq:lse_convolution_def}
\end{equation}
Therefore, the LSE of the subsurface velocity field can be obtained by a convolution of the kernel $l_{ij}$ and the surface quantities $E_j$ over the horizontal plane $S$.
As a result, the vertical coordinate $z$ is decoupled, and~\eqref{eq:lse_convolution_def} can be evaluated independently for each $x$-$y$ plane. 

For efficiently computing $l_{ij}$, we apply the convolution theorem to~\eqref{eq:lse_convolution_def} and transform it into pointwise multiplications in the Fourier wavenumber space. Similarly,~\eqref{eq:lse_optimize} can also be expressed in the convolutional form. This yields
\begin{align}
    \hat{\tilde{u}}_i &= \hat{l}_{ij}\hat{E}_j, \\
    \left\langle{\hat{E}_m^* \hat{E_j}}\right\rangle \hat{l}_{ij} &= \left\langle{\hat{E}_m^* \hat{u}_i}\right\rangle, \label{eq:lse_optimization_wn}
\end{align}
where $\hat{f}$ denotes the Fourier coefficients of a quantity $f$ and $\hat{f}^*$ denotes the complex conjugate of $\hat{f}$.
The above equations can be solved independently for each wavenumber in the horizontal directions, each vertical location and each velocity component.

\subsection{\label{sec:dataset}Dataset descriptions}
\begin{figure}
    \centering
    \includegraphics{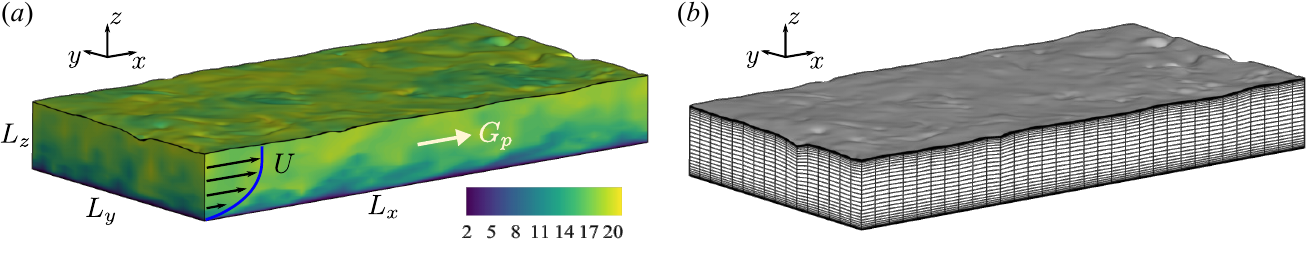}
    \caption{\label{fig:dns_sketch}(a) Set-up of the turbulent open-channel flow. The contours illustrate the instantaneous streamwise velocity $u$ of one snapshot from the simulation. (b) Sketch of the boundary-fitted curvilinear coordinate system. For illustration purposes, the domain is stretched, and the grid resolution is lowered in the plots.}
\end{figure}

The reconstruction models proposed above are trained and tested with datasets obtained from the DNS of a turbulent open-channel flow.
Figure~\ref{fig:dns_sketch}(\sublabel{a}) shows the configuration of the turbulent open-channel flow, which is doubly periodic in the streamwise ($x$) and spanwise ($y$) directions. In the vertical ($z$) direction, the bottom wall satisfies the no-slip condition, and the top surface satisfies the free-surface kinematic and dynamic boundary conditions. The utilised simulation parameters are listed in table~\ref{tab:simulation_param}. The domain size is $L_x\times L_y \times L_z = 4\pi h \times 2\pi h \times h$, with $L_z$ being the mean depth of the channel. The flow is driven by a constant mean pressure gradient $G_p$ in the streamwise direction. The friction Reynolds number is defined as ${Re}_\tau=u_\tau h/\nu=180$, where $u_\tau=\sqrt{G_p/\rho}$ is the friction velocity, $h$ is the mean depth of the channel, and $\nu$ is the kinematic viscosity. Flows with different Froude numbers, which are defined based on the friction velocity as $Fr_\tau = u_\tau/\sqrt{gh}$, are considered. \rev{The domain size $4\pi h \times 2\pi h \times h$ is sufficiently large to obtain domain-independent one-point turbulence statistics in open-channel flows at ${Re}_\tau=180$~\citep{wang2020}. The autocorrelations studied by~\citet{pinelli2022} indicate that a spanwise domain size of $6h$ is sufficiently large for the spanwise scales to be decorrelated; in the streamwise direction, a much larger domain length, $L_x>24h$, is needed for complete decorrelation. Owing to the higher computational cost of the deformable free-surface simulations compared to the rigid-lid simulations, here we use a smaller domain length $L_x=4\pi h$. The normalised autocorrelation of the streamwise velocity fluctuations, $R_u(x', z) = \langle u(x') u(x+ x') \rangle/\langle u^2\rangle$, where $\langle \cdot\rangle$ denotes the plane average, is smaller than $0.065$ at the largest separation distance $L_x/2$ across the depth of the channel, which is small enough to allow most streamwise structures to be captured, consistent with the numerical result of~\citet{pinelli2022}.} 

\rev{\subsubsection{Simulation method and parameters}}
\rev{To simulate free-surface motions, we discretise the governing equations on a time-dependent boundary-fitted curvilinear coordinate system, $\boldsymbol{\xi} = (\xi_1, \xi_2, \xi_3)$, which is defined as~\citep{xuan2019},
\begin{equation}
\xi_1 = x, \quad \xi_2 = y, \quad \xi_3 = \frac{z}{\eta(x,y,t)+h}h. \tag{\theequation \sublabel{a}--\sublabel{c}}
\end{equation}
The above transformation maps the vertical coordinate $z\in [0, \eta+h]$ to the computational coordinate $\xi_3\in [0, h]$. The mass and momentum equations are written using the curvilinear coordinates as
\begin{subequations}\label{eq:govern_eqns}
    \begin{align}
        \frac{\p (J^{-1} u_i)}{\p t} - \frac{\p (J^{-1} U^j_g u_i)}{\p \xi^j}
+ \frac{\p (J^{-1} U^j u_i)}{\p \xi^j} = 
 & -\frac{\p}{\p \xi^j}\left(J^{-1} \frac{\p \xi^j}{\p x_i} p\right)
+ \frac{1}{\Rey} \frac{\p}{\p \xi^j}\left({J^{-1} g^{ij} \frac{\p u_i}{\p \xi^j}}\right),\label{eq:mom_curv} \\
\frac{\p U^j}{\p \xi^j} &= 0,\label{eq:conti_curv}
    \end{align}
\end{subequations}
where $J=\det \left( \p \xi^i / \p x_j \right)$ is the Jacobian determinant of the transformation; $U^j=u_k (\p \xi^j / \p x_k)$ is the contravariant velocity; $U^j_g= \dot{x}_k({\p \xi^j}/{\p x_k})$ is the contravariant velocity of the grid where $\dot{x}_k(\xi_j)$ is the moving velocity of $\xi_j$ in the Cartesian coordinates;  $g^{ij}=(\p \xi^i / \p x_k)(\p \xi^j / \p x_k)$ is the metric tensor of the transformation. The above curvilinear coordinate system, which is illustrated in figure~\ref{fig:dns_sketch}(\sublabel{b}), enables the discretised system to more accurately resolve the surface boundary layers.}

\rev{The solver utilises a Fourier pseudo-spectral method for discretisations in the horizontal directions, $\xi_1$ and $\xi_2$, and a second-order finite-difference scheme in the vertical direction $\xi_3$. The temporal evolution of \eqref{eq:govern_eqns} is coupled with the evolution of the free surface $\eta(x, y, t)$; the latter is determined by the free-surface kinematic boundary condition, written as
\begin{equation}
    \eta_t = w - u \eta_x - v \eta_y.
\end{equation}
The above equation is integrated using a two-stage Runge--Kutta scheme. In each stage after the $\eta$ is updated,~\eqref{eq:govern_eqns} is solved to obtain the velocity in the domain bounded by the updated water surface.
}
This simulation method has been validated with various canonical wave flows~\citep{yang2011a,xuan2019} and has been applied to several studies on turbulent free-surface flows, such as research on the interaction of isotropic turbulence with a free surface~\citep{guo2010} and the turbulence--wave interaction~\citep*{xuan2020,xuan2022}. More details of the numerical schemes and validations are described in~\citet{xuan2019}.

\begin{figure}
    \centering
    \includegraphics{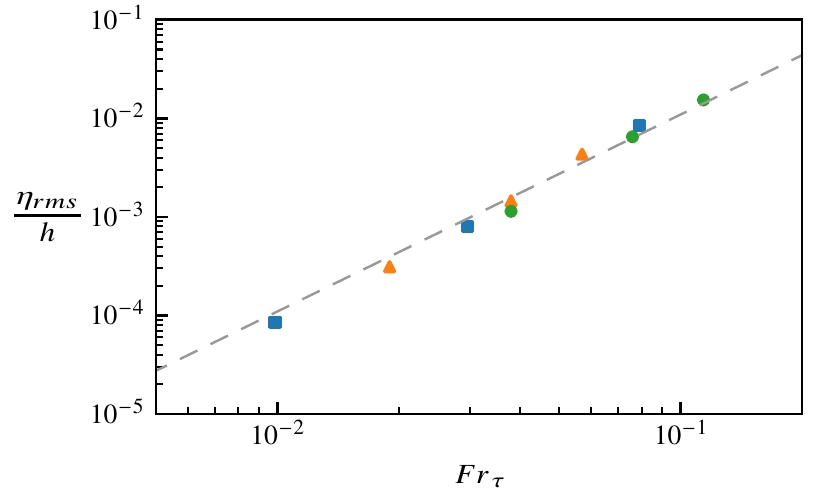}
    \caption{\label{fig:comparison_etarms}\rev{Comparison of the root-mean-square free-surface fluctuations with literature: the present DNS results (\fullsquare) and the DNS results of~\cite{yoshimura2020} (\fullcirc) and~\citet{yokojima2002} (\fulltriangle) are compared. The dashed line (\broken) is an approximated parameterization of $\eta_{rms}/h$ proposed by~\citet{yoshimura2020}; i.e. $\eta_{rms}/h= 4.5\times 10^{-3}{Fr}^2$ where $Fr=\overline{U}/\sqrt{gh}$ is defined based on the bulk mean velocity $\overline{U}$ and, in the present study, $\overline{U}=15.6 u_\tau$.}}
\end{figure}

\rev{
\begin{table}
    \begin{center}
    \begin{tabular}{cccccccc}
        $\displaystyle Fr_\tau = \frac{u_\tau}{\sqrt{gh}}$  &  $\displaystyle Re_\tau = \frac{u_\tau h}{\nu}$  & $L_x \times L_y \times L_z$  & $N_x\times N_y \times N_z$  & \rev{$\Delta x^+$} & \rev{$\Delta y^+$} & \rev{$\Delta z^+_{min}$}  &  \rev{$\Delta z^+_{max}$} \\
        $0.08$  & $180$ & $4\pi h \times 2\pi h \times h$ & $256\times 256\times 128$ & \rev{$8.8$} & \rev{$4.4$} & \rev{$0.068$} & \rev{$2.7$} \\
        $0.03$  & $180$ & $4\pi h \times 2\pi h \times h$ & $256\times 256\times 128$ & \rev{$8.8$} & \rev{$4.4$} & \rev{$0.068$} & \rev{$2.7$} \\
        $0.01$  & $180$ & $4\pi h \times 2\pi h \times h$ & $256\times 256\times 128$ & \rev{$8.8$} & \rev{$4.4$} & \rev{$0.068$} & \rev{$2.7$}
    \end{tabular}
    \caption{\label{tab:simulation_param}The simulation parameters of the turbulent open-channel flows. The superscript ${}^+$ denotes the quantities normalised by wall units $\nu/u_\tau$.}
    \end{center}
\end{table}

The horizontal grid, following the Fourier collocation points, is uniform. Near the bottom of the channel and the free surface, the grid is clustered with a minimum vertical spacing $\Delta^+_{min}=0.068$, where the superscript ${}^+$ denotes the length normalised by wall units $\nu/u_\tau$; the maximum vertical spacing at the center of the channel is $\Delta z^{+}_{max}=2.7$. The above grid resolutions, as listed in the last four columns in table~\ref{tab:simulation_param}, are comparable to those used in the DNS of turbulent rigid-lid open-channel flows~\citep{yao2022} and turbulent free-surface open-channel flows~\citep{yoshimura2020}. Figure~\ref{fig:comparison_etarms} compares the free-surface fluctuation magnitudes in the present simulations with the literature~\citep{yokojima2002,yoshimura2020}. We find that the present results agree well with their data. The profiles of the mean velocity and Reynolds stresses are also in agreement with the results reported by~\citet{yoshimura2020} (not plotted).
}

\subsubsection{\label{sec:preprocess}Dataset preprocessing and training}
For each case, the dataset consists of a series of instantaneous flow snapshots stored at constant intervals for a period of time. The intervals and the total number of samples collected, denoted by $\Delta t u_\tau/h$ and $N$, respectively, are listed in table~\ref{tab:simulation_param}. We note that the simulations are performed on a computational grid of size $N_x \times N_y \times N_z=256\times 256 \times 128$. The simulation data are interpolated onto the input and output grids, which are uniform grids with coarser resolutions, as listed in table~\ref{tab:nn_param}, to reduce the computational cost of the training process and the consumption of GPU memory. \rev{Although the reduced resolutions remove some fine scale structures from the training data compared to the original DNS data, we find that an increased resolution has a negligible effect on the reconstruction (see details in \S\,3 of the supplementary material), indicating that the present resolution is adequate for resolving the reconstructed structures.} We apply random translations to the flow fields in the horizontal periodic directions, i.e. in the $x$ and $y$ directions, and obtain a total of $N_i$ samples, as listed in the last column of table~\ref{tab:simulation_param}. This process, known as data augmentation, which artificially increases the amount of data via simple transformations, can help reduce overfitting and increase the generalization performance of NNs, especially when it is difficult to obtain big data~\citep{shorten2019}. Here, owing to the high computational cost of free-surface flow simulations, we adopt this technique to expand the datasets. \rev{As discussed in~\S\,\ref{sec:cnn}, this process can also improve the translation invariance of the CNN~\citep{kauderer-abrams2017}.} 
\rev{The surface elevation in the input is normalised by $\eta_{rms}$ of each case; the surface velocities are normalised by the root-mean-square value of all three velocity components, ${\langle (u_s^2 + v_s^2 + w_s^2)/3 \rangle}^{1/2}$; the subsurface velocities are normalised by the root-mean-square magnitude of velocity fluctuations in the subsurface flow field, ${\langle (u'^2+v'^2+w'^2)/3 \rangle}^{1/2}$. This normalisation uniformly scales three velocity components because it is necessary to obtain a divergence-free velocity. }

Each dataset is split into training, validation and test sets with $75\%$, $10\%$ and $15\%$ of the total snapshots, respectively. The training set is used to train the network and the validation set is used to monitor the performance of the network during the training process. The training procedure is stopped when the loss function~\eqref{eq:loss} computed over the validation set no longer improves, an indication that the model is overfitting \rev{(see \S\,5 of the supplementary material for learning curves)}. The test set is used to evaluate the network. It should be noted that the training, validation and test sets are obtained from different time segments of the simulations such that the flow snapshots in the test set do not have strong correlations with those in the training set.

\rev{The LSE also uses the same training set as in CNN to determine the operator $\mathcal{L}$, i.e. the LSE also uses the interpolated simulation data for reconstructions. However, random translations are not applied to the dataset for the LSE reconstructions because the LSE method is inherently translation-invariant.}

\begin{table}
    \begin{center}
\rev{
    \begin{tabular}{ccccccc}
        $\displaystyle Fr_\tau$  & $\Delta t u_\tau/h$  &  $N$  &  $N_i$  & $N_\text{train}$  &  $N_\text{val}$ & $N_\text{test}$ \\
        $0.08$  &  $0.01$  &  $4,061$   &  $16,244$ & $12,183$ & $1,624$ & $2,437$ \\
        $0.03$  &  $0.01$  &  $4,964$   &  $19,056$ & $14,292$ & $1,905$ & $2,859$ \\
        $0.01$  &  $0.01$  &  $4,986$   &  $19,944$ & $14,958$ & $1,994$ & $2,992$
    \end{tabular}
}
    \caption{\label{tab:dataset}\rev{The sampling parameters and the sizes of the datasets, including: the non-dimensionalised sampling interval $\Delta t u_\tau/h$, the total number of snapshots sampled from simulations $N$, the total number of snapshots after augmentation $N_i$, the number of snapshots in the training set $N_\text{train}$, the number of snapshots in the validation set $N_{\text{val}}$, and the number of snapshots in the test set $N_\text{test}$.}}
    \end{center}
\end{table}

\section{\label{sec:result}Flow reconstruction performance}
In this section, the performance of the CNN and LSE models in terms of reconstructing turbulent free-surface open-channel flows is compared both qualitatively and quantitatively. The subsurface flow structure features that can be inferred by the two models are also discussed. \rev{It should be noted that in this section, the cases with different ${Fr}_\tau$ are processed individually, i.e.\ each case has its own kernel weights for the CNN or linear operator for the LSE such that the reconstruction errors are minimized for that specific case.} 

\subsection{\label{sec:instant}Instantaneous flow features}
We first inspect the instantaneous flow fields to qualitatively assess the performance of the CNN and LSE methods.
In this section, unless specified otherwise, we use one snapshot in the test set for the case with ${Fr}_\tau=0.08$ as an example to compare the reconstructed flow field and the ground truth from DNS\@. \rev{It should also be noted that hereafter, unless specified otherwise, the DNS results as the ground truth refer to the velocity fields on the uniform reconstruction grid as described in~\S\,\ref{sec:preprocess}.} 

\begin{figure}
    \includegraphics[width=0.99\textwidth]{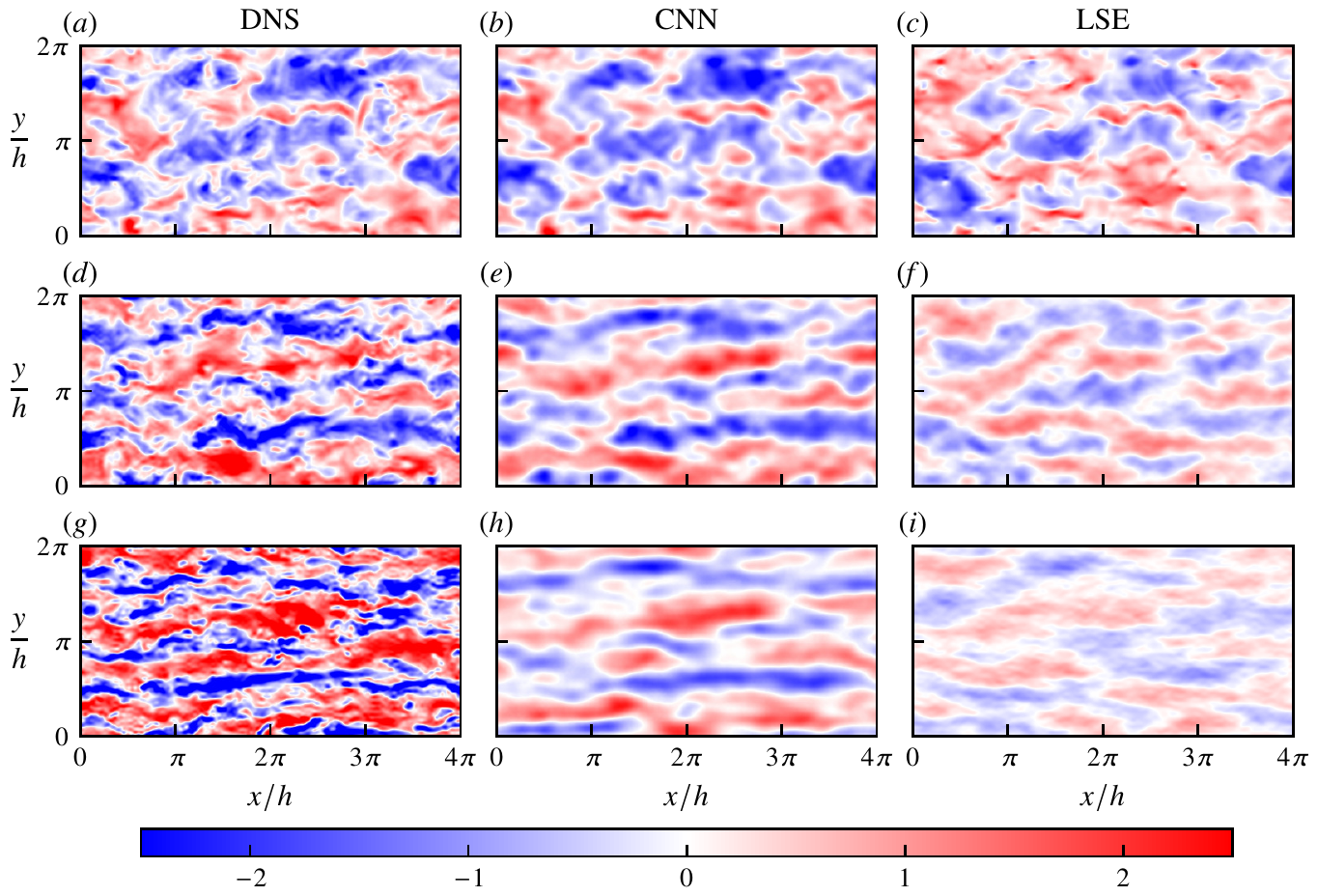}
    \caption{\label{fig:comparison_velocity_plane_u}Comparisons among the instantaneous streamwise velocity fluctuations $u'$ (\sublabel{a},\sublabel{d},\sublabel{g}) obtained from the DNS results and the fields reconstructed by (\sublabel{b},\sublabel{e},\sublabel{h}) the CNN and (\sublabel{c},\sublabel{f},\sublabel{i}) the LSE methods. The $x$-$y$ planes at (\sublabel{a}--\sublabel{c}) $z/h=0.9$, (\sublabel{d}--\sublabel{f}) $z/h=0.6$ and (\sublabel{g}--\sublabel{i}) $z/h=0.3$ are plotted for the case of ${Fr}_\tau=0.08$.}
\end{figure}

Figure~\ref{fig:comparison_velocity_plane_u} compares the subsurface streamwise velocity fluctuations estimated by the CNN and LSE methods with the DNS results. Near the surface at $z/h=0.9$, the DNS result (figure~\ref{fig:comparison_velocity_plane_u}\sublabel{a}) shows that the regions with negative $u'$ values, i.e.\ the regions with low-speed streamwise velocities, are characterised by patchy shapes. The high-speed regions distributed around the low-speed patches can appear in narrow band-like patterns. The patchy low-speed and streaky high-speed structures of the near-surface streamwise motions are consistent with the open-channel flow simulation results obtained by~\citet{yamamoto2011}. Despite some minor differences, both the CNN and LSE methods provide reasonably accurate reconstructions. In the high-speed regions, the LSE method (figure~\ref{fig:comparison_velocity_plane_u}\sublabel{c}) predicts the amplitudes of the velocity fluctuations slightly more accurately than the CNN method (figure~\ref{fig:comparison_velocity_plane_u}\sublabel{b}). Moreover, the contours of the LSE reconstruction, especially in high-speed regions, have sharper edges, indicating that the LSE method can predict higher gradients than the CNN method. This result suggests that the LSE approach can capture more small-scale motions, thereby yielding slightly better reconstruction performance in the high-speed regions with narrow-band patterns. On the other hand, the CNN method can more accurately predict the low-speed patches, whereas the LSE method tends to underestimate the fluctuations in the low-speed regions. Overall, both methods perform similarly well in the near-surface regions.

Away from the free surface, close to the centre of the channel ($z/h=0.6$) (figure~\ref{fig:comparison_velocity_plane_u}\sublabel{d}), we can observe increases in the streamwise scales of $u'$ compared to those of the near-surface region ($z/h=0.9$). The high-speed and low-speed regions start to appear as relatively long streamwise streaks and alternate in the spanwise direction. At this height, the reconstructions provided by both the CNN and LSE methods are not as accurate as their reconstructions near the surface. However, the CNN method (figure~\ref{fig:comparison_velocity_plane_u}\sublabel{e}) provides a better reconstruction than the LSE method (figure~\ref{fig:comparison_velocity_plane_u}\sublabel{f}) in that the former better captures the streaky pattern of $u'$. The structures predicted by the LSE method generally appear shorter than those in the ground truth. Furthermore, although the reconstructed velocities provided by both methods exhibit underpredicted magnitudes, the velocity magnitudes predicted by the CNN method are clearly higher than those predicted by the LSE method and follow the DNS results more closely.

Farther away from the surface at $z/h=0.3$, the streamwise streaks become more pronounced, and the intensities of the fluctuations become stronger (figure~\ref{fig:comparison_velocity_plane_u}\sublabel{g}); these are the typical features of turbulent flows affected by strong shear near the wall. The CNN method (figure~\ref{fig:comparison_velocity_plane_u}\sublabel{h}), despite its underestimated velocity magnitudes and missing small-scale structures, can still predict distributions for the large-scale streaks that are similar to the DNS result. In contrast, the streaky structures in the reconstruction provided by LSE (figure~\ref{fig:comparison_velocity_plane_u}\sublabel{i}) are far less discernable, and the magnitudes of the velocity fluctuations are also weaker.

The results obtained for the vertical and spanwise velocity fluctuations are presented in appendix~\ref{sec:instantaneous_vw}. The performance difference between the CNN and LSE methods is similar to what we have observed for the streamwise velocity component.

\begin{figure}
    (\sublabel{a})\\
    \includegraphics[width=0.99\textwidth,trim=0.02in 0 0 0.03in,clip]{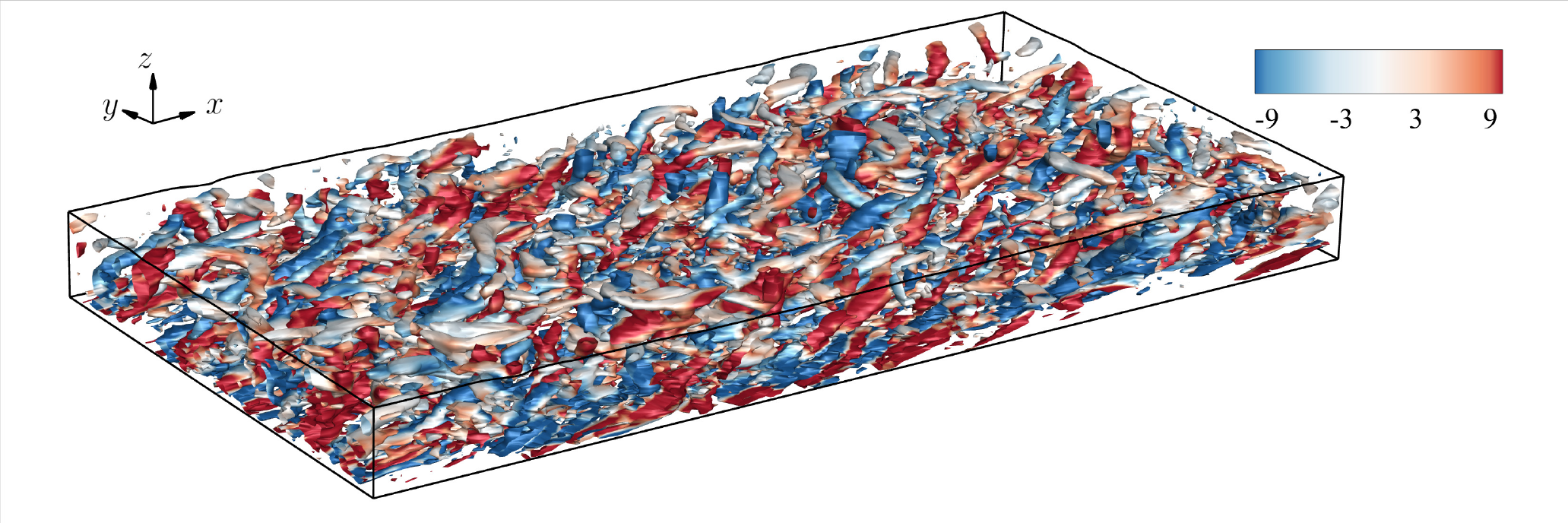}\\
    (\sublabel{b})\\
    \includegraphics[width=0.99\textwidth,trim=0.02in 0 0 0.00in,clip]{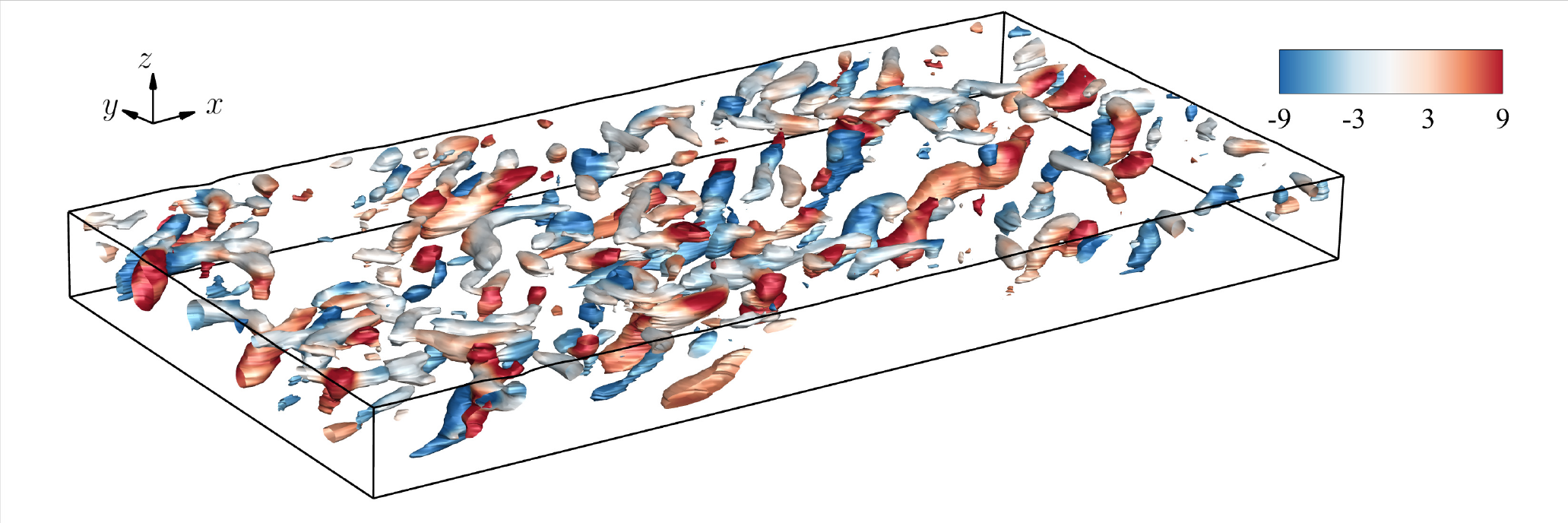}\\
    (\sublabel{c})\\
    \includegraphics[width=0.99\textwidth,trim=0.02in 0 0 0.03in,clip]{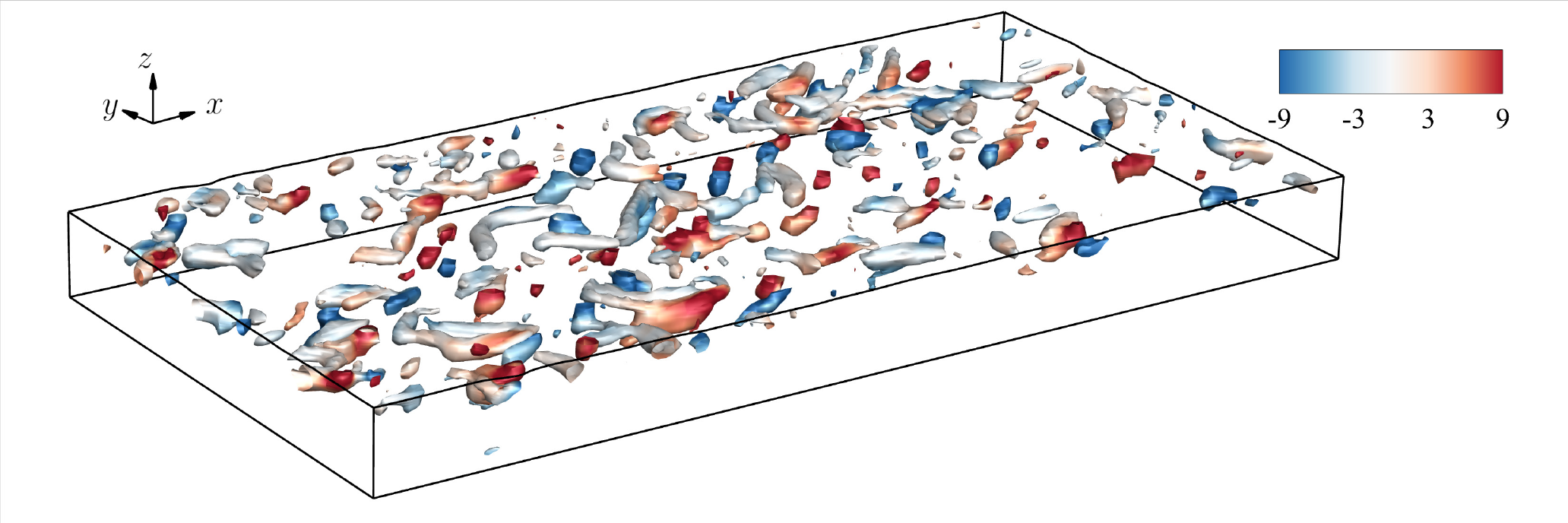}\\
    \caption{\label{fig:vortex}Comparisons of the instantaneous vortex structures among the (\sublabel{a}) DNS, (\sublabel{b}) CNN reconstruction and (\sublabel{c}) LSE reconstruction. The vortex structures are educed with the criterion $\lambda_2<0$~\citep{jeong1995}. The isosurface with $0.9\%$ of the minimum $\lambda_2$ value is plotted and is coloured by $\omega_z$.}
\end{figure}

Based on the above observations, we find that the CNN method can reconstruct the characteristic turbulent coherent structures away from the surface, e.g.\ the streamwise streaks near the wall, to some extent. To further assess the capabilities of the reconstruction methods regarding instantaneous flow structures, we extract vortices from the DNS data and the reconstruction data using the $\lambda_2$-criterion~\citep{jeong1995}. It should be noted that both the CNN and LSE models are optimised to minimize the errors in the velocities and do not guarantee similarity in the vortical structures. Considering the imperfect velocity reconstruction results observed above, we are interested in whether the two models can reconstruct physically reasonable vortical structures.

\begin{figure}
    (\sublabel{a})\\
    \includegraphics[width=0.99\textwidth,trim=0.02in 0.3in 0 0.2in,clip]{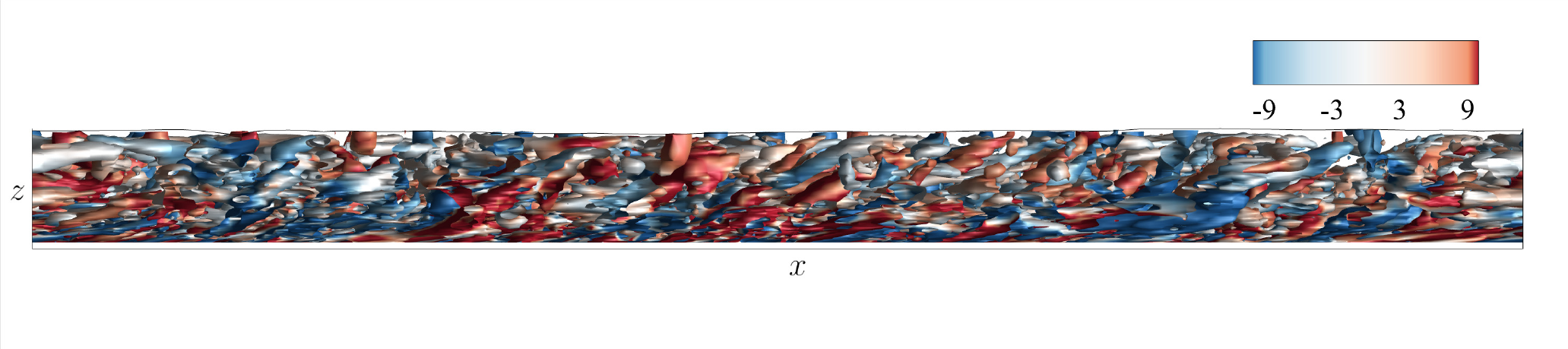}
    (\sublabel{b})\\
    \includegraphics[width=0.99\textwidth,trim=0.02in 0.2in 0 0.1in,clip]{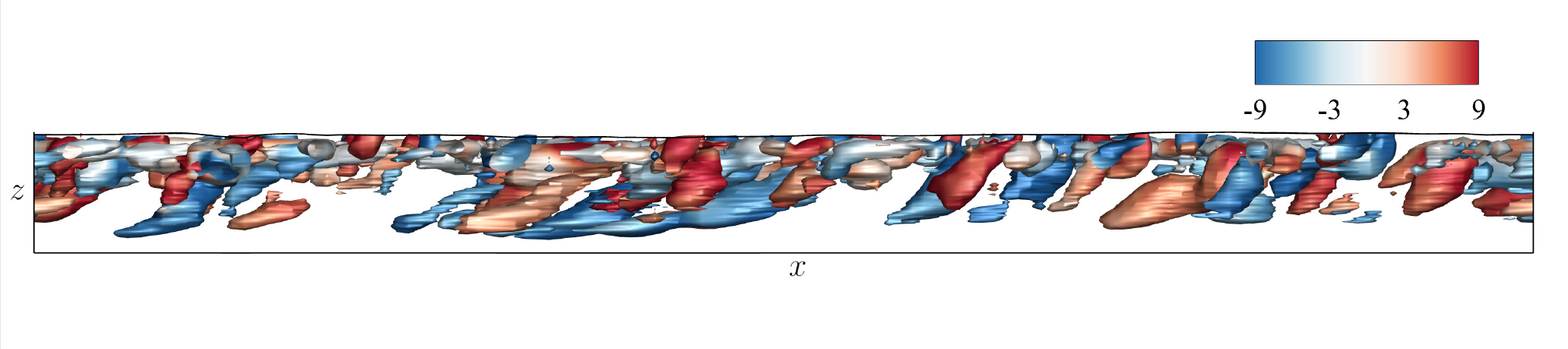}
    (\sublabel{c})\\
    \includegraphics[width=0.99\textwidth,trim=0.02in 0.2in 0 0.1in,clip]{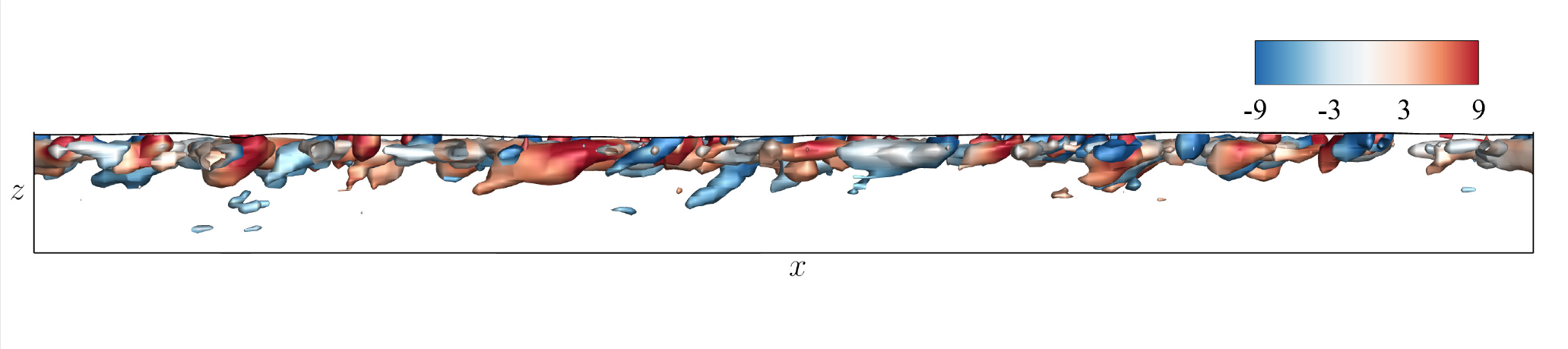}
    \caption{\label{fig:vortex_xz}Side-views of the inclined vortices in the (\sublabel{a}) DNS, (\sublabel{b}) CNN reconstruction and (\sublabel{c}) LSE reconstruction, as plotted in figures~\ref{fig:vortex}(\sublabel{a}), \ref{fig:vortex}(\sublabel{b}) and~\ref{fig:vortex}(\sublabel{c}), respectively. The vortex structures are educed by the $\lambda_2$-criterion and the isosurfaces are coloured by $\omega_z$.}
\end{figure}

As shown in figures~\ref{fig:vortex} and~\ref{fig:vortex_xz}, compared to the DNS results, the reconstructed flow fields contain much fewer vortical structures. Specifically, most small-scale vortices and near-wall vortices are not captured by the reconstructions. This is another indication that the two reconstruction methods have difficulties in capturing small-scale structures. However, we note that the vortical structures recovered by the CNN method extend farther away from the surface than the vortices recovered by the LSE method. Most vortices in the LSE reconstruction are located near the surface. This difference indicates that the CNN method is more capable than the LSE method of reconstructing the flow structures away from the free surface. Moreover, the vortices that span the entire channel, which can be seen more clearly in the side view plotted in figure~\ref{fig:vortex_xz}(\sublabel{b}), are inclined towards the flow direction ($+x$-direction), similar to what can be observed in the DNS result (figure~\ref{fig:vortex_xz}\sublabel{a}). The inclination angles of these vortices, approximately $45^\circ$, are consistent with the features of the rolled-up vortices that originate from the hairpin vortices in the shear-dominated near-wall region~\citep{christensen2001}. This result suggests that the vortical structures reconstructed by the CNN method are indeed turbulent coherent structures that would be expected in a turbulent open-channel flow.

It is evident from our inspection of the instantaneous velocity fields and vortical structures that the CNN method has a notable advantage over the traditional LSE method in reconstructing subsurface flows, especially away from the free surface. We find that the CNN method captures the characteristic features of the near-wall turbulence more accurately than the LSE method.

\subsection{\label{sec:stats}\rev{Normalised mean squared errors of reconstruction}}
\begin{figure}
    \centering
    \includegraphics[width=0.99\textwidth]{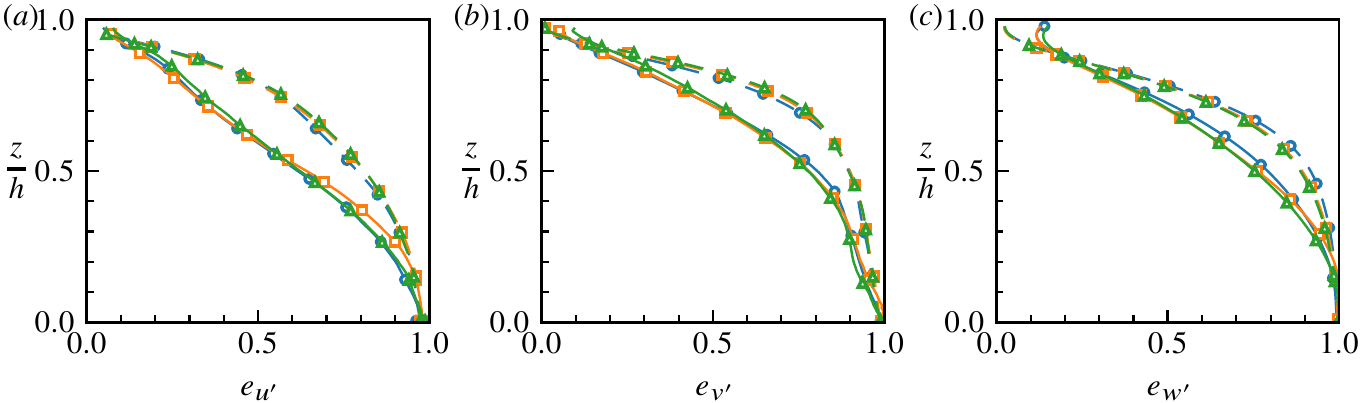}
    \caption{\label{fig:error_l2}\rev{Normalised mean squared errors \eqref{eq:l2_error_def} of the CNN ({\full}) and LSE ({\broken}) reconstructions at $Fr_\tau=0.08$ (\textcolor{colorC0}{\opencirc}), $Fr_\tau=0.03$ (\textcolor{colorC1}{\opensquare}) and $Fr_\tau=0.01$ (\textcolor{colorC2}{\opentri}) for the (\sublabel{a}) streamwise, (\sublabel{b}) spanwise and (\sublabel{c}) vertical velocity fluctuations between the DNS and reconstructed fields.}}
\end{figure}

\rev{To quantify the performance of the CNN and LSE methods, we compute a normalised mean squared error defined as
\begin{equation}
    e_i(z) = \frac{\left\langle \iint_{x,y} {\left| \tilde{u}_i - u_i \right|}^2 \,\mathrm{d}x\mathrm{d}y \right\rangle}{\left\langle \iint_{x,y} u_i^2 \,\mathrm{d}x\mathrm{d}y \right\rangle}. \label{eq:l2_error_def}
\end{equation}
which measures the reconstruction error of each velocity component on each horizontal plane. We note that the performance evaluation is based on the test set of the corresponding case, i.e.\ the averaging $\langle\cdot\rangle$ in~\eqref{eq:l2_error_def} is conducted on the snapshots in the test set.
The reconstruction errors of the CNN and LSE models for different cases are plotted in figure~\ref{fig:error_l2}. We find that the reconstruction performances for cases with different $Fr$ are comparable  and, therefore, here we focus on the differences between the CNN and LSE reconstructions.}

\rev{As shown in figure~\ref{fig:error_l2}, the overall reconstruction accuracy of the CNN reconstructions are significantly higher than the accuracy of the LSE method. Near the free surface, both the CNN and LSE methods have low reconstruction errors, indicating that the reconstruction of the near-surface flow is relatively more accurate. The LSE even slightly outperforms the CNN in predicting the vertical velocity fluctuations. Farther away from the free surface, the errors of the LSE increase more rapidly than the errors of the CNN. As a result, the advantage of the CNN method becomes more pronounced near the centre of the channel. For example, at $z/h=0.6$, the reconstruction of $u'$ provided by the LSE method has a normalised mean squared error of $e_{u'}=0.68$, which is $45\%$ larger than the CNN reconstruction result of $e_{u'}=0.47$.
This result is also consistent with our observation of the instantaneous flow fields (figure~\ref{fig:comparison_velocity_plane_u}\sublabel{d}--\sublabel{f}).}  

\rev{Both the CNN and LSE reconstructions become less accurate away from the free surface, suggesting that the flow field away from the free surface is physically more difficult to predict. This behaviour is because the flow structures away from the surface have fewer direct effects on the free-surface motions and, therefore, become increasingly decorrelated with the free-surface motions. Such an accuracy decrease when reconstructing flows away from where the measurements are taken was also observed by~\citet{guastoni2021} and~\citet{wang2022} for wall-bounded flows when these authors used the CNN method and adjoint-variational method, respectively, to predict turbulent flows based on wall measurements.
However, it should be noted that the mean squared error defined in~\eqref{eq:l2_error_def} is an aggregated measure of the reconstruction accuracy for the different structures in a flow. As discovered in~\S\,\ref{sec:instant} above, the CNN reconstructions of large-scale structures near the bottom wall still have a decent similarity with the ground truth despite that the small-scale motions are missing. In other words, the loss of the reconstruction accuracy varies for structures at different scales, and, therefore, the scale-specific reconstruction performance is investigated in~\S\,\ref{sec:spectral} below.
}

\subsection{\label{sec:spectral}Spectral properties of reconstruction}
\rev{In this section, we further evaluate the CNN and LSE reconstruction performance with respect to spatial scales based on the Fourier spectrum of the reconstructed velocities. Here, the coefficients of the Fourier modes of the velocity $u_i$, which are denoted by $\hat{u}_i$, are computed with respect to the streamwise wavenumber $k_x$ and spanwise wavenumber $k_y$, thereby representing the turbulence fluctuations at different streamwise scales and spanwise scales, respectively. A two-dimensional spectrum with respect to ($k_x$, $k_y$) can also be computed. But to conduct easier comparisons, here, we are interested only in the one-dimensional spectrum.}

\begin{figure}
    \centering
    \includegraphics{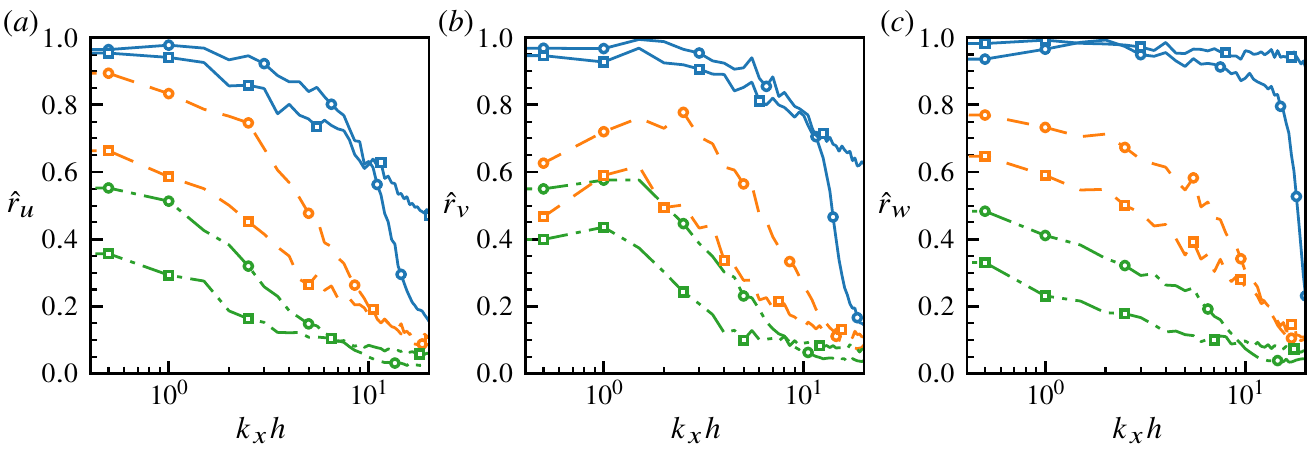}
    \includegraphics{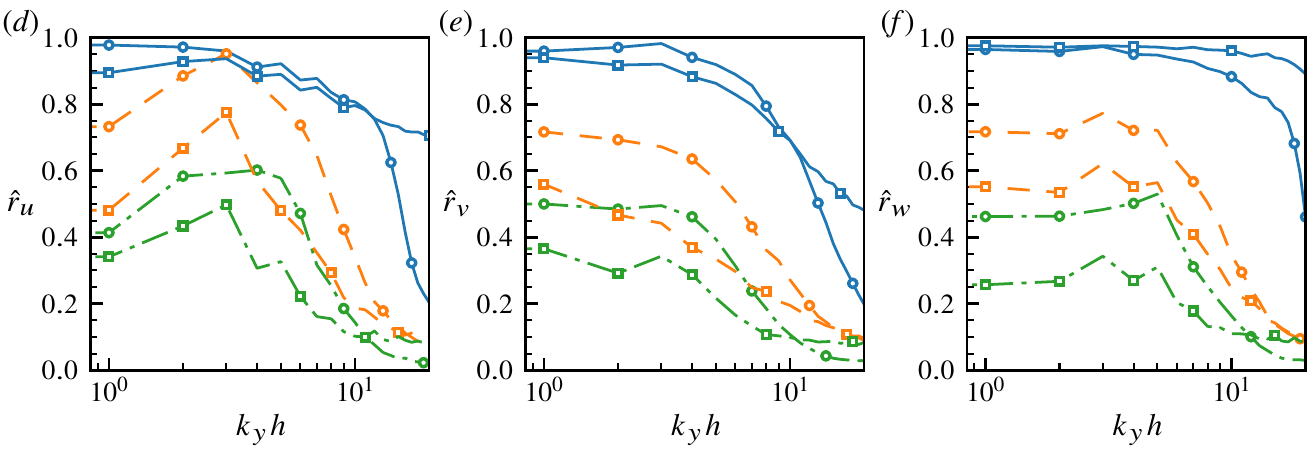}
    \caption{\label{fig:spectra_corr}\rev{Relative amplitudes of the Fourier coefficients of reconstructed velocity fluctuations compared to the ground truth: (\textit{a,d}) $\hat{r}_u$, (\textit{b,e}) $\hat{r}_v$ and (\textit{c,f}) $\hat{r}_w$, for the streamwise, spanwise and vertical velocity fluctuations, respectively. The spectra are evaluated at $z/h=0.9$ (\textcolor{colorC0}{\full}), $z/h=0.6$ (\textcolor{colorC1}{\broken}) and $z/h=0.3$ (\textcolor{colorC1}{\chain}) for the CNN reconstructions (\opencirc) and LSE reconstructions (\opensquare). The first row (\textit{a}--\textit{c}) and the second row (\textit{d}--\textit{f}) plot the spectra with respect to the streamwise wavenumber $k_x$ and the spanwise wavenumber $k_y$, respectively.}}
\end{figure}

\rev{We first assess the reconstruction accuracy in terms of the amplitude of the spectrum by computing the relative Fourier amplitude, $\hat{r}_i = |\hat{\tilde{u}}_i|/|\hat{u}_i|$, which is the ratio of the amplitude of the reconstructed mode $|\hat{\tilde{u}}_i|$ to the ground truth $|\hat{u}_i|$. Because the Fourier amplitude spectrum is the square root of the energy spectrum, the relative amplitude spectrum can also be viewed as a measure of how much turbulent kinetic energy at each scale is recovered by the reconstructions.

Figure~\ref{fig:spectra_corr} plots the relative amplitude spectrum $\hat{r}_i$ for several representative vertical locations of the case with $Fr_\tau=0.08$. In general, for each scale, $\hat{r}_i$ of either the CNN and LSE methods decreases with the increasing depth, which is consistent with the conclusion drawn above from the mean squared errors. Regarding the scale-wise performance, we observe that $\hat{r}_i$ at small wavenumbers are generally higher than $\hat{r}_i$ at high wavenumbers, indicating that large-scale motions can be predicted more accurately than small-scale motions.

Near the surface at $z/h=0.9$, the CNN method outperforms the LSE method in predicting the streamwise and spanwise velocity fluctuations at small wavenumbers, yet the LSE still has good accuracy and has a slightly better performance than CNN for vertical velocity fluctuations. At larger wavenumbers, approximately $k_x h>10$ and $k_y h>10$, the performance of the CNN method drops rapidly and becomes worse than that of the LSE method. This result indicates that although the presented CNN model can accurately predict near-surface flow motions over a wide range of scales, it has difficulty reconstructing small-scale motions. This result is also consistent with our observation regarding the instantaneous streamwise velocity fields (figures~\ref{fig:comparison_velocity_plane_u}\sublabel{a}--\sublabel{c}) where the CNN method more accurately predicts the velocities in the low-speed regions, which appear more frequently in patchy shapes that have large scales, while the LSE method can better reconstruct the narrow-band high-speed regions that are more skewed towards small scales.

Away from the surface at $z/h=0.6$ and $z/h=0.3$, the CNN method outperforms the LSE method at almost all scales, and its performance lead is even more pronounced for large-scale motions. For example, at $z/h=0.6$, the relative amplitudes of the CNN reconstructions at $k_x h < 3$ and $k_y h < 3$ for the streamwise, spanwise and vertical velocity components are at least $38\%$, $24\%$ and $19\%$ higher than the relative amplitudes of the LSE reconstructions, respectively.

Figure~\ref{fig:spectra_corr}(\sublabel{d}) shows an interesting phenomenon concerning the spanwise structures of the streamwise velocity fluctuations. The relative amplitude $\hat{r}_u$ for the CNN reconstructions exhibits a peak around $k_y h \approx 3$, indicating that the CNN-reconstructed streamwise velocity around this spanwise scale agrees with the DNS result better than at other scales. Moreover, the relative amplitude around $k_y h \approx 3$, being greater than $0.9$ at $z/h=0.6$ and still as high as $0.6$ at $z/h=0.3$, decreases only slowly with the increasing depth (not plotted for other examined depths). This behaviour indicates that the corresponding spanwise structures in the streamwise velocity fluctuations can be reconstructed by the CNN with good accuracy across the channel depth. This result quantitatively supports our observation that the CNN method can accurately capture the large-scale streaky pattern exhibited by the streamwise velocity fluctuations that alternate in signs in the spanwise direction (see figure~\ref{fig:comparison_velocity_plane_u}\sublabel{e,h}). By comparison, the performance of the LSE method at this scale is not as good, which is consistent with the lack of streaky patterns in the reconstructed streamwise velocity fields shown in figure~\ref{fig:comparison_velocity_plane_u}(\sublabel{f,i}).
}

\begin{figure}
    \centering
    \includegraphics{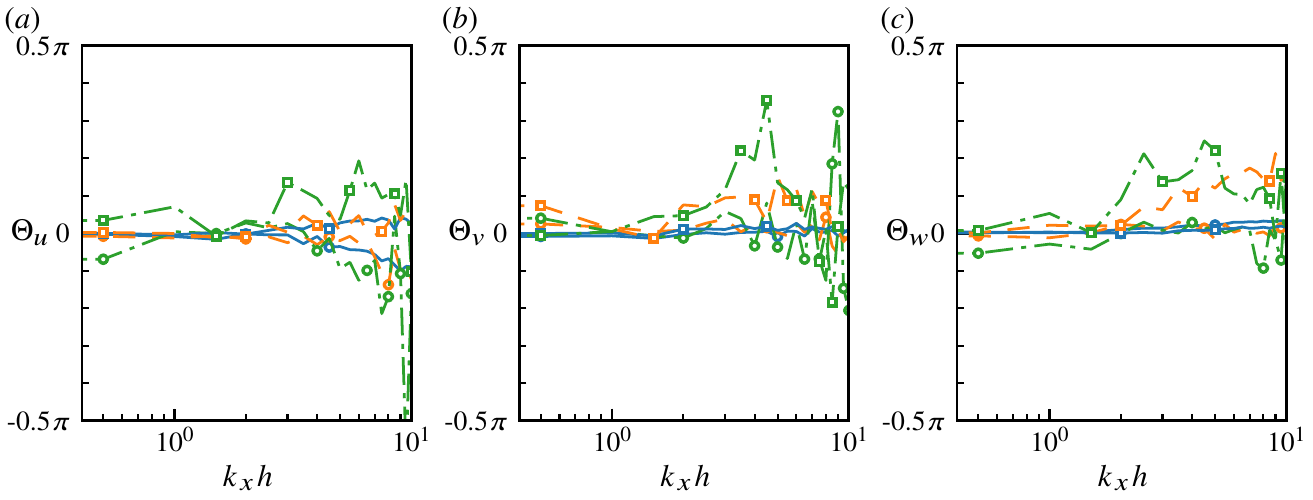}
    \includegraphics{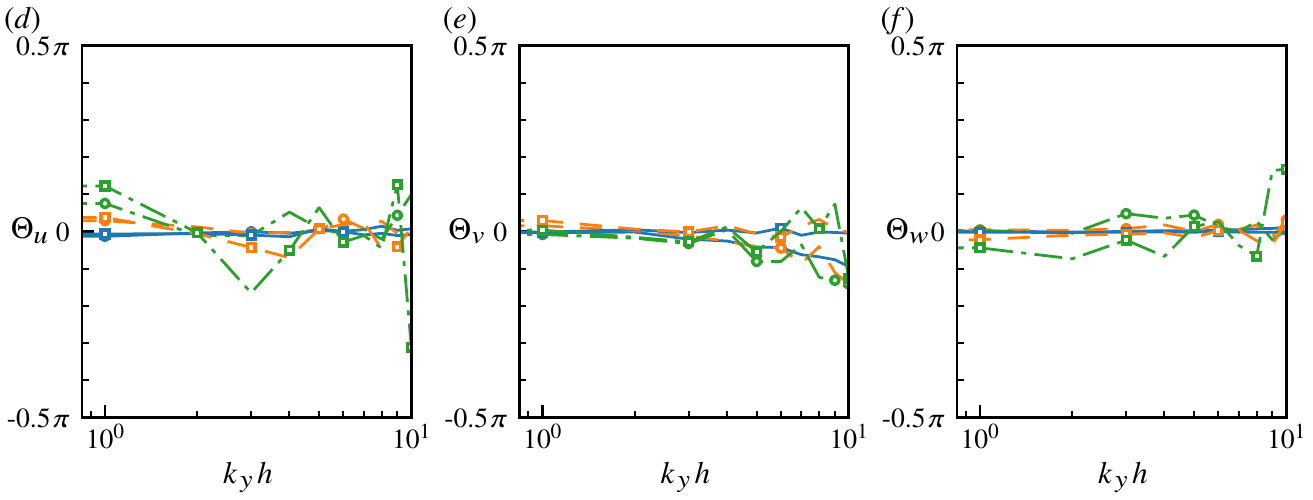}
    \caption{\label{fig:spectra_corr_angle}\rev{Mean phase errors of the reconstructed velocity fluctuations~\eqref{eq:def_phase_error}: (\textit{a,d}) $\hat{\Theta}_u$, (\textit{b,e}) $\hat{\Theta}_v$ and (\textit{c,f}) $\hat{\Theta}_w$ for the streamwise, spanwise and vertical velocity fluctuations, respectively. The phase errors are evaluated at $z/h=0.9$ (\textcolor{colorC0}{\full}), $z/h=0.6$ (\textcolor{colorC1}{\broken}) and $z/h=0.3$ (\textcolor{colorC2}{\chain}) for the CNN reconstructions (\opencirc) and LSE reconstructions (\opensquare). The first row (\textit{a}--\textit{c}) and the second row (\textit{d}--\textit{f}) plot the errors with respect to the streamwise wavenumber $k_x$ and the spanwise wavenumber $k_y$, respectively.}}
\end{figure}

\rev{
Next, we measure the phase errors of the reconstructions. Denoting $\theta$ as the phase angle of a Fourier mode, a Fourier coefficient can be expressed as $\hat{u}_i = |\hat{u}_i|e^{i\theta}$. The phase difference between the reconstruction $\hat{\tilde{u}}_i$ and the ground truth $\hat{u}_i$ can then be calculated as $\Delta \theta =\tilde{\theta} - \theta =\mathrm{arg}(\hat{\tilde{u}}_i/\hat{{u}}_i) = \mathrm{arg}(e^{i\Delta\theta})$, where $\mathrm{arg}(\cdot)$ denotes the angle of a complex value. A large phase difference means that the predicted structure is shifted by a large distance relative to the true structure. To assess the phase errors for all snapshots in the dataset, we calculate a mean phase difference between the reconstructions and the DNS results defined as
\begin{equation}
    \Theta_i = \mathrm{arg}\left(\frac{\left\langle |\hat{u}_i| e^{i\Delta\theta} \right\rangle}{\left\langle |\hat{u}_i| \right\rangle}\right). \label{eq:def_phase_error}
\end{equation}
In the above equation, the phase difference is weighted by the amplitude of the fluctuations such that the phase errors of stronger modes have more weights in the mean value.

The results of the mean phase errors are plotted in figure~\ref{fig:spectra_corr_angle}. It should be noted that because both the CNN and LSE methods cannot reconstruct small-scale motions with consistent accuracy, we focus only on the phase errors at low wavenumbers. It is found that in the wavenumber range $k_x h < 6$ and $k_y h< 6$, the phase errors of the CNN reconstructions are within $7.8^{\circ}$ at $z/h=0.9$ and $z/h=0.6$, and are within $16.9^{\circ}$ at $z/h=0.3$, indicating that the CNN reconstructions have low phase errors. Furthermore, when we compare the full error $|\hat{u}_i - \hat{\tilde{u}}_i|$ that accounts for both the amplitude and phase difference to the amplitude difference $|\hat{u}_i (1-\hat{r}_i)|$ at $z/h=0.3$ in the above wavenumber range $k_x h < 6$ and $k_y h< 6$, we find that the amplitude error is close to the full error, indicating that the amplitude error is dominant. In other words, although the CNN model under-estimates the fluctuating amplitudes of these large-scale structures, it can predict the phases of these fluctuation modes with relatively good accuracy. By comparison, the phase errors of the LSE reconstructions are significantly larger in the above wavenumber range, reaching $31^{\circ}$ at $z/h=0.6$ and exceeding $63^{\circ}$ at $z/h=0.3$. We also note that the LSE produces considerably larger phase errors in $u'$ at $k_y h \approx 3$, corresponding to the streaky patterns, than the CNN.}

\rev{The above quantitative assessments further confirm that the CNN method is more accurate than the LSE method in predicting subsurface flows from free-surface measurements. For turbulence motions across a wide range of scales, the fluctuating amplitudes and phases predicted by the CNN method are more accurate than the reconstructions provided by the LSE method. In particular, we find that the low-wavenumber spanwise structures, which correspond to streamwise streaks characterised by positive--negative $u'$ values alternating in the spanwise direction, are effectively captured by the CNN method.}

The structures captured by the reconstructions also depict how subsurface turbulence influences free-surface motions. Because the reconstructions are essentially mappings from the surface quantities to the subsurface velocities~\eqref{eq:mapping}, the structures that can be captured must influence the surface motions, either by directly impacting the surface or by correlating with the near-surface structures that have manifestations on the surface. This enables us to gain some understanding of the interactions between subsurface turbulence and the free surface. The CNN can capture large-scale streamwise streaks better than other near-wall structures, indicating that these streaks have strong relations with free-surface motions. Considering that the near-wall streaks are closely related to hairpin vortices that originate from the near-wall shear layer, we can infer that the correlations between the streamwise streaks and the free-surface motions are due to the evolving hairpin vortices that impact the free surface; this is consistent with the findings by~\citet{nagaosa2003} and~\citet{sanjou2011} that the hairpin vortices can roll up from the wall and rise to the free surface. The relations between the rolled-up hairpin vortices and the surface motions are supported by the presence of tilted vortices in the flow field reconstructed by the CNN method (figure~\ref{fig:vortex_xz}). Therefore, we see that the CNN reconstruction method can serve as a promising tool for revealing the dynamics of free-surface turbulence. In the next section, how the free-surface dynamics affect the reconstruction models is discussed further.

\section{\label{sec:discussions}Discussion}
\subsection{\label{sec:difference_nn_lse}Differences between the CNN and LSE methods}
The results presented in the preceding sections show that the CNN method is more accurate than the LSE method. As both methods treat each snapshot from the given dataset as an independent sample, the reconstructions are based only on the spatial relations between the surface quantities and the subsurface flow field. Therefore, the higher accuracy of the CNN method indicates that it can map some complex surface--subsurface correlations that the LSE model cannot describe. In this section, the differences between the methodologies of the two methods are discussed to provide insights into why the CNN method is more successful than the LSE method in estimating the mappings from the surface quantities to the subsurface flow field.

First, although both the LSE and CNN methods are mainly based on convolution operations, the two methods differ in how their convolution operations are structured and the sizes of the kernels used in their convolutions. As described by the alternative form of the LSE method in~\eqref{eq:lse_convolution_def}, the linear transformation $\mathcal{L}$ of the LSE method is equivalent to convolving the surface inputs with a set of kernels in the $x$- and $y$-directions and then linearly superimposing the convolutions from different surface quantities. With four quantities available on the surface ($E_j$, where $j=1,2,3,4$, i.e. $u_s$, $v_s$, $w_s$, $\eta$) and three output components for the subsurface velocity ($\tilde{u}_i$, where $i=1,2,3$), twelve pairs of convolution operations exist. For each convolution operation, the convolution kernel is periodic in the $x$ and $y$ directions and is as large as the whole domain. In other words, the LSE method uses twelve whole-field convolutions as the mapping between the surface signatures and subsurface flow structures, which means that these twelve kernels in the LSE method need to describe all the characteristic features related to the surface--subsurface interactions to produce accurate reconstructions.

In contrast to the LSE method, which employs whole-field kernels that process the entire domain at once, the CNN method uses small kernels with two to five elements in each dimension, as listed in table~\ref{tab:nn_param}. Their convolution with an input can be considered a process of extracting or constructing certain local features over the entire domain by moving the kernel window across the input. To capture more types of features, one can either adjust the number and sizes of the kernels in each convolutional layer or stack multiple convolutional layers together. \rev{Although each kernel in the CNN method has a limited size, combined with the downsampling or upsampling that change the number of grid points in feature maps, the kernels in different layers can process features at different scales efficiently.}

Another crucial difference between the LSE and CNN methods concerns their ability to describe nonlinear relations. The transformation in the LSE method is completely linear. Although the convolution operations in the CNN method are also linear, nonlinear activation functions are added among its layers to introduce nonlinear effects. It should be noted that without the nonlinear layers, a multi-layer convolutional network can still be expressed as one linear transformation. The nonlinear layers, combined with the layer stacking, allow the CNN to create complex mappings between the surface and subsurface features. Various studies have discovered that the ability of NNs to process nonlinear dynamics often enables them to achieve better performance than linear methods when applied to turbulent flow problems (e.g. see~\citealp{brunton2020} for a review).

\rev{Considering the above differences between the CNN and LSE methods, we believe that the CNN method's advantage over the LSE method can be attributed to the former's ability to capture nonlinear relations between the free-surface motions and the subsurface turbulence field. Moreover, we note that the reconstruction accuracy of the LSE method drops faster than the CNN method when moving away from the free surface (figure~\ref{fig:error_l2}), suggesting that the nonlinearity is more necessary for reconstructing structures away from the free surface than near the surface. An example is the streamwise streaks near the bottom and the associated vortical structures, which are captured by the CNN but not by the LSE method. This result suggests that nonlinear dynamics may be active in the process of the streaks and the vortices rolled up from the streaks influencing the free-surface motions. The surface--subsurface relations in the reconstruction models are further analysed in the next section.}

\subsection{Interpreting subsurface flow reconstructions}
In this section, we aim to gain further understanding of how the LSE and CNN methods utilise surface information to infer the subsurface velocity and how such information is related to the dynamics of free-surface turbulence.

\subsubsection{Convolution kernels of the LSE method\label{sec:lse_kernel}}
As discussed in the preceding sections, the LSE method depends on the convolution kernels $l_{ij}$ to reconstruct the subsurface velocities from the surface inputs. When substituting $E_j=\delta(\boldsymbol{x})$ into~\eqref{eq:lse_convolution_def}, we obtain
\begin{equation}
    \tilde{\boldsymbol{u}}_i(\boldsymbol{x}) = l_{ij}(\boldsymbol{x}),
\end{equation}
which means that each kernel $l_{ij}$ is the least-squares approximated response of $\tilde{\boldsymbol{u}}_i$ to a point impulse of $\boldsymbol{E}_j$ on the surface. Therefore, by looking at the convolution kernels, we can gain a better understanding of the surface--subsurface correlations that the LSE utilises to predict the subsurface velocities.

\begin{figure}
    \includegraphics{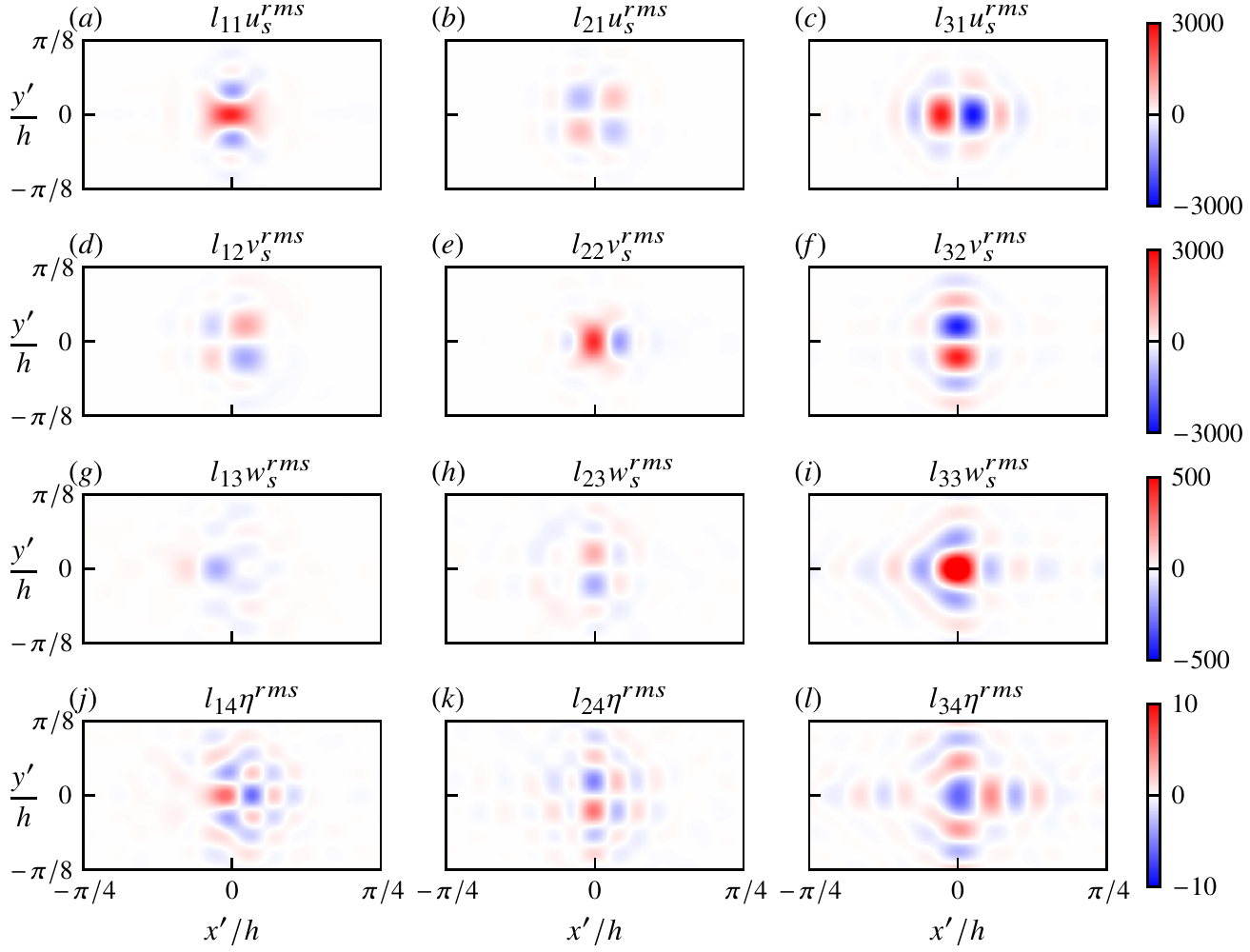}
    \caption{\label{fig:lse_kernel_f08_09}Contours of the scaled kernels, $l_{ij}E_j^{rms}$, for the LSE prediction of the velocity at $z/h=0.9$ of the flow with ${Fr}_\tau=0.08$. Each row shows the kernels of one surface variable $E_j$: (\sublabel{a}--\sublabel{c}) $u_s$, (\sublabel{d}--\sublabel{f}) $v_s$, (\sublabel{g}--\sublabel{i}) $w_s$ and (\sublabel{j}--\sublabel{l}) $\eta$. Each column corresponds to one component of the predicted subsurface velocity: (\sublabel{a},\sublabel{d},\sublabel{g},\sublabel{j}) $\tilde{u}'$, (\sublabel{b},\sublabel{e},\sublabel{h},\sublabel{k}) $\tilde{v}'$ and (\sublabel{c},\sublabel{f},\sublabel{i},\sublabel{l}) $\tilde{w}'$.}
\end{figure}

Figure~\ref{fig:lse_kernel_f08_09} shows the kernels utilised for predicting the velocity at a near-surface plane ($z/h=0.9$) in the flow with ${Fr}_\tau=0.08$. This plane is investigated because the LSE method achieves good accuracy only near the surface. Considering that the reconstructions are the linear superpositions of the products of $l_{ij}$ and $E_j$ (see~\ref{eq:lse_convolution_def}), to reflect the relative contributions of each surface variable to the final reconstructions, the kernels plotted in figure~\ref{fig:lse_kernel_f08_09} are pre-multiplied by the root-mean-square values of their corresponding surface variables.
Figure~\ref{fig:lse_kernel_f08_09} indicates that both $u_s$ and $v_s$ are correlated with the prominent subsurface structures. In contrast, the kernels associated with $\eta$ and $w_s$ have smaller amplitudes, indicating that the surface elevation and vertical fluctuations have weaker influences on the subsurface flows predicted by the LSE method. Next, we focus on analysing the structures associated with $u_s$ and $v_s$.

\begin{figure}
    \includegraphics{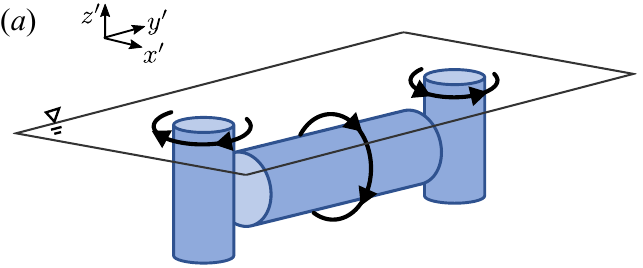}
    \includegraphics{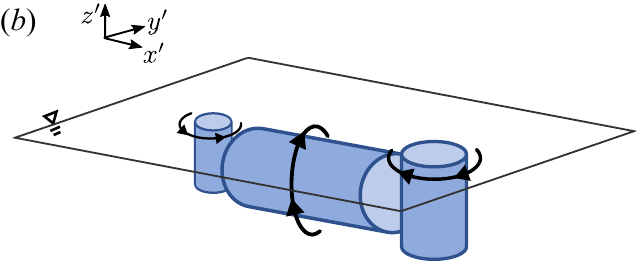}
    \caption{\label{fig:lse_vortex_f08}Illustrations of the vortical structures associated with the LSE kernels (\sublabel{a}) $l_{i2}$ and (\sublabel{b}) $l_{i3}$, respectively, i.e.\ the LSE predicted flow structures induced by $u_s$ and $v_s$, respectively.}
\end{figure}

The velocities induced by $u_s$ (figure~\ref{fig:lse_kernel_f08_09}\sublabel{a}--\sublabel{c}) correspond to vortical motions, as illustrated in figure~\ref{fig:lse_vortex_f08}(\sublabel{a}). Underneath $u_s$ is a vortex that rotates in the spanwise direction with a positive $\omega_y$; this vortex brings fluids upward on its $-x'$ side and downward on its $+x'$ side, which corresponds to positive and negative $w'$ values on its $+x'$ and $-x'$ sides, respectively (figure~\ref{fig:lse_kernel_f08_09}\sublabel{c}). On the $-y'$ and $+y'$ sides, two vertical vortices rotate in opposite directions, with $\omega_z<0$ on the $-y'$ side and $\omega_z>0$ on the $+y'$ side. The vertically rotating motions result in the four-quadrant pattern of the spanwise velocity fluctuations (figure~\ref{fig:lse_kernel_f08_09}\sublabel{b}) and the backward flow in the streamwise velocity fluctuations (figure~\ref{fig:lse_kernel_f08_09}\sublabel{a}).

\begin{figure}
    \includegraphics[width=5.2in, trim=0.03in 0 0 0.03in, clip]{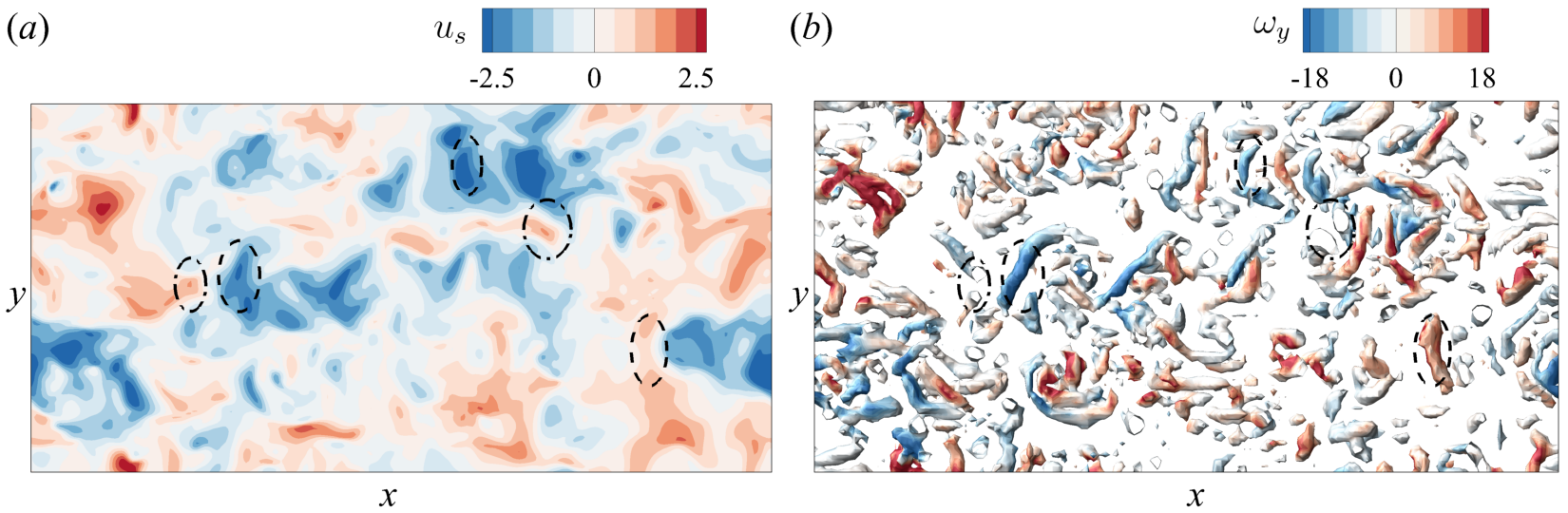}
    \caption{\label{fig:lse_vortex_instant_u}Relationships between (\sublabel{a}) $u_s$ and (\sublabel{b}) the subsurface vortical structures, which are educed by the $\lambda_2$ criterion. In (b), only the near-surface vortices above $z/h=0.8$ are plotted, and the vortices are coloured by their spanwise vorticity $\omega_y$. Dashed circles and dash-dotted circles mark the $u_s$ examples induced by the spanwise vortices and vertical vortices, respectively. Note that the vertical vortices observed in the top view appear as hollowed circles.}
\end{figure}

When we inspect the instantaneous flow field of $u_s$ and the subsurface vortices obtained from DNS, as shown in figures~\ref{fig:lse_vortex_instant_u}(\sublabel{a}) and~\ref{fig:lse_vortex_instant_u}(\sublabel{b}), respectively, we find multiple flow regions that satisfy the correlations between $u_s$ and the subsurface vortices as described above. In figure~\ref{fig:lse_vortex_instant_u}, some typical spanwise vortex-induced $u_s$ motions and vertical vortex-induced $u_s$ motions are highlighted by the dashed circles and dash-dotted circles, respectively. This result indicates that our interpretations of the flow structures based on the LSE kernel match the real flows.

\begin{figure}
    \includegraphics[width=5.2in, trim=0.03in 0 0 0.03in, clip]{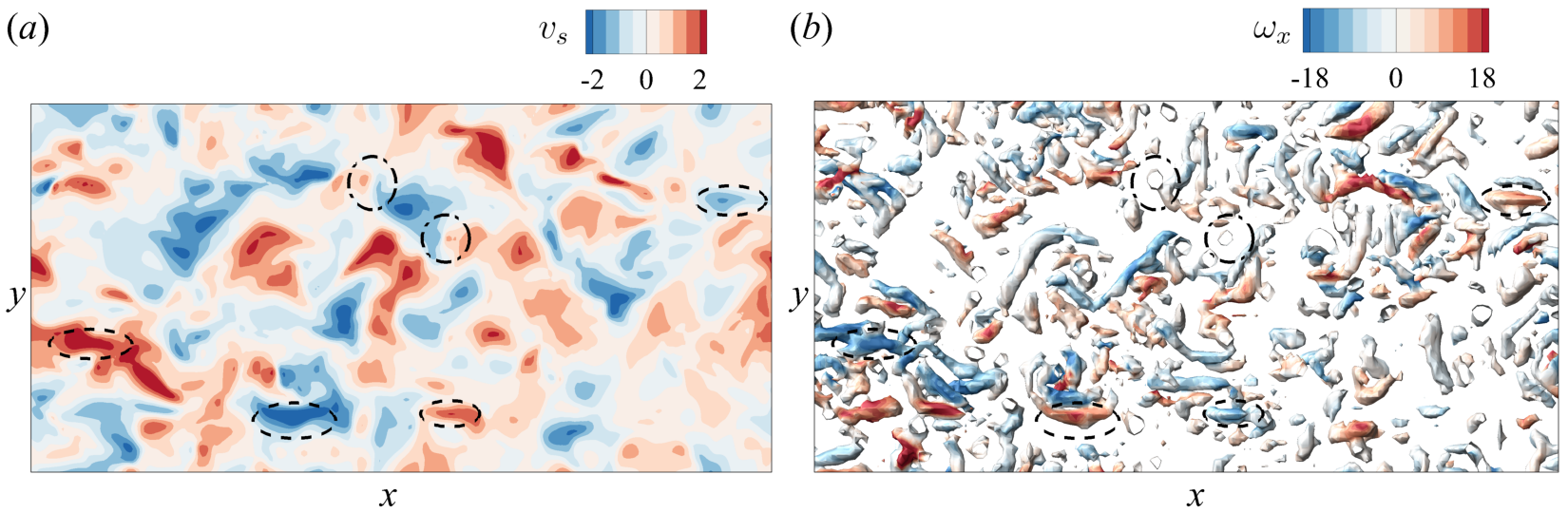}
    \caption{\label{fig:lse_vortex_instant_v}Relationships between (\sublabel{a}) $v_s$ and (\sublabel{b}) the subsurface vortical structures, which are educed by the $\lambda_2$ criterion. In (b), only the near-surface vortices above $z/h=0.8$ are plotted, and the vortices are coloured by their streamwise vorticity $\omega_x$. Dashed circles and dash-dotted circles mark the $v_s$ examples induced by the streamwise vortices and vertical vortices, respectively. Note that the vertical vortices observed in the top view appear as hollowed circles.}
\end{figure}

The vortical structures associated with $v_s$ are illustrated in figure~\ref{fig:lse_vortex_f08}(\sublabel{b}). At the centre, a streamwise vortex with a negative $\omega_x$ leads to a positive $w'$ on the $-y'$ side and a negative $w'$ on the $+y'$ side (figure~\ref{fig:lse_kernel_f08_09}\sublabel{f}). The vertically rotating vortices on the $+x'$ and $-x'$ sides have negative $\omega_z$ and positive $\omega_z$ values, respectively, corresponding to the four-quadrant pattern in $u'$ (figure~\ref{fig:lse_kernel_f08_09}\sublabel{d}) and the negative--positive--negative pattern in $v'$ (figure~\ref{fig:lse_kernel_f08_09}\sublabel{e}). However, we note that unlike the velocity induced by $u_s$ (figure~\ref{fig:lse_kernel_f08_09}\sublabel{e}), the four-quadrant distribution of $u'$ is asymmetric, with the vortex on the $+x'$ side being stronger. This asymmetric distribution indicates that a vertical vortex is more likely to be found downstream of $v_s$.
In the instantaneous flow field (figure~\ref{fig:lse_vortex_instant_v}), we can find several locations where the streamwise vortices and vertical vortices coincide with $v_s$ at the surface in a way that is consistent with the above description, confirming that $v_s$ is affected by both types of vortices. Although the streamwise and vertical vortices do not necessarily appear together, some of the streamwise vortices are partially connected to the free surface; i.e. the streamwise vortices become vertical in the surface-connected regions. Vortex connections are important phenomena when vortices interacting with a free surface (\citealp{shen1999}; \citealp*{zhang1999a}). In the instantaneous flow field, we find that the connections often occur near the downstream ends of the streamwise vortices; this finding is consistent with the asymmetric pattern of the vortical structures shown in figure~\ref{fig:lse_vortex_f08}, i.e. a vertical vortex on the $+x'$ side of the streamwise vortex. We note that the other vertical vortex on the $-x'$ side is not obvious in the instantaneous flow field because it is hidden by the superimposed structures. That is, in the LSE method, the structures induced by the $v_s$ values at different surface locations are superimposed to yield the reconstructions, and weaker structures may be superseded by stronger structures.

Finally, we note that most instantaneous near-surface vortices do not appear exactly as the structures shown in figure~\ref{fig:lse_vortex_f08} \rev{because the reconstruction is the convolution of the LSE kernels with the four surface variables (see~\eqref{eq:lse_convolution_def}), with $u_s$ and $v_s$ being the dominant ones as discussed above. Moreover, we note that the determination of the LSE kernels~\eqref{eq:lse_optimize} has the information of the two-point correlations of the surface variables and, therefore, the obtained kernels consider not only the correlations between the surface motions and the subsurface velocities, but also the expectancy of the surrounding surface structure given a surface value. As a result, the reconstructed structures are not isolated but should be considered as superposition of the key structures shown in figure~\ref{fig:lse_vortex_f08} according to the surface velocity distribution, which can lead to the merging and cancellation among structures. For example, most horizontal vortices do not strictly align with the $x$- or $y$-direction and may be curved. These inclined horizontal vortices can be considered as adding streamwise and spanwise vortices together. Cancellations of structures can also occur. For example, we observe that spanwise vortices and vertical vortices may not appear together in the instantaneous flow fields as in figure~\ref{fig:lse_vortex_f08}(\sublabel{a}). This can be due to the side vertical vortices cancelled by the surrounding structures because the two-point correlation between $u_s$ and $v_s$ suggests that a positive $u_s$ is correlated with a negative $v_s$ region on its $+y'$ side and a positive $v_s$ region on its $-y'$ side slightly in front it; such velocity distribution means that there exists a possibility that the side vortices associated with $u_s$ may be cancelled by the vertical vortex on the $-x'$ side of $v_s$.}

To summarize, we find that the representative near-surface vortices can be described by their linear correlations with $u_s$ and $v_s$, which allows the LSE method to extract them into its kernels through least-squares approximations and then use them for reconstructions. \rev{However, such linear correlations break down when moving away from the surface, leading to rapid increases in the errors of LSE reconstructions. Moreover, the cores of the spanwise or streamwise vortices derived from the LSE kernels using the $\lambda_2$ criterion (not plotted) are located at $z/h=0.92$, further supporting that the horizontal vortical structures described by the LSE appear only near the surface. This is also consistent with the observation that the LSE method cannot capture vortical structures outside the near-surface region (see figures~\ref{fig:vortex} and~\ref{fig:vortex_xz}).} By comparison, the CNN model, which captures the nonlinear relationships between the free-surface motions and the subsurface flow structures, utilises the free-surface information in a different way than the linear LSE model does, which is analysed below in~\S\,\ref{sec:saliency_map}.

\subsubsection{Saliency map of the CNN method\label{sec:saliency_map}}
Owing to the complex structures and nonlinearity of CNNs, understanding how CNNs work has always been a challenge, and the interpretability of CNNs is an active research area~\citep[see, e.g.][]{zhang2021}. For example, the kernels in a CNN model are stacked and used with nonlinear activation functions. As a result, the actual flow structures that a kernel represents may depend on the features in all previous layers, which makes the analysis in~\S\,\ref{sec:lse_kernel} unsuitable for the CNN. Therefore, in the present work, we do not aim for a complete understanding of the CNN model. Instead, we focus on seeking indications of which surface quantities and what surface features may be most important for subsurface flow reconstruction. Such information may provide insights into surface--subsurface interaction dynamics and the design of surface measurements.

We employ saliency maps to measure the importance of each surface input to the subsurface flow reconstruction. In image classification applications, a saliency map is a way to quantify the importance levels of different pixels in an image to the identification of a particular class~\citep{simonyan2014}. Assuming that the input and output of an NN are denoted by $\boldsymbol{I}$ and $\boldsymbol{S}_c$, respectively, the network is expressed as 
\begin{equation}
    \boldsymbol{S}_c(I) \approx \boldsymbol{w}^T \boldsymbol{I} + \boldsymbol{b},
\end{equation}
where $\boldsymbol{w}$ denotes the weights applied to the different components and grid points of the input and $\boldsymbol{b}$ is the bias. Because a CNN consists of multiple layers and is nonlinear, both $\boldsymbol{w}$ and $\boldsymbol{b}$ depend on $\boldsymbol{I}$. For a specific input $\boldsymbol{I}_0$, we can estimate the weight as
\begin{equation}
    \boldsymbol{w} \approx \left.\frac{\partial \boldsymbol{S}_c}{\partial \boldsymbol{I}}\right|_{\boldsymbol{I}_0}.
\end{equation}
The saliency map, denoted by $G$, is defined as the absolute value of $\boldsymbol{w}$, i.e. $G=|\boldsymbol{w}|$. According to this definition, a pixel with a larger saliency value means that the output of the NN is more sensitive to \rev{the changes in the value at this point} than its surroundings. \rev{In other words, the features in high salient regions are important to the objective that the CNN is optimised for because the variations of these features can affect the network output more.}

To understand the importance of the surface inputs to the flow reconstruction, we use a saliency map to investigate the sensitivity of the encoder output to the input. Note that the output of the encoder layer has significantly fewer dimensions than the input; therefore, substantial information is discarded by the encoder. The optimisation of the NN should enable the encoder to retain the most important information. As a result, a saliency map evaluated on the encoder part can indicate what surface quantities the encoder pays most attention to, i.e. what surface features are most important for the CNN when predicting subsurface flows. The output of the encoder consists of multiple components, and we compute saliency maps for each component and superimpose them to obtain a single saliency map representing all the salient features extracted by the encoder. \rev{The saliency maps are often noisy with grid-scale (or pixel-scale) oscillations, which may be smoothed by taking local averages of the gradient~\citep{smilkov2017}. \citet{smilkov2017} proposed to obtain the averaged gradients from a set of inputs with stochastic noises added. Here, considering that the underlying physics should be translation invariant, we simply use translations to smooth out grid-scale oscillations, i.e. apply periodic translations to the input, shift the obtained gradients back, and then take the averages to obtain the saliency map. We find that shifting the input in the range of $-3\sim 3$ grid points in both the $x$- and $y$-directions is enough to eliminate the grid-scale noises; further increasing the shifting range yields no essential changes in the obtained saliency maps.}

\begin{figure}
    \centering
    \includegraphics{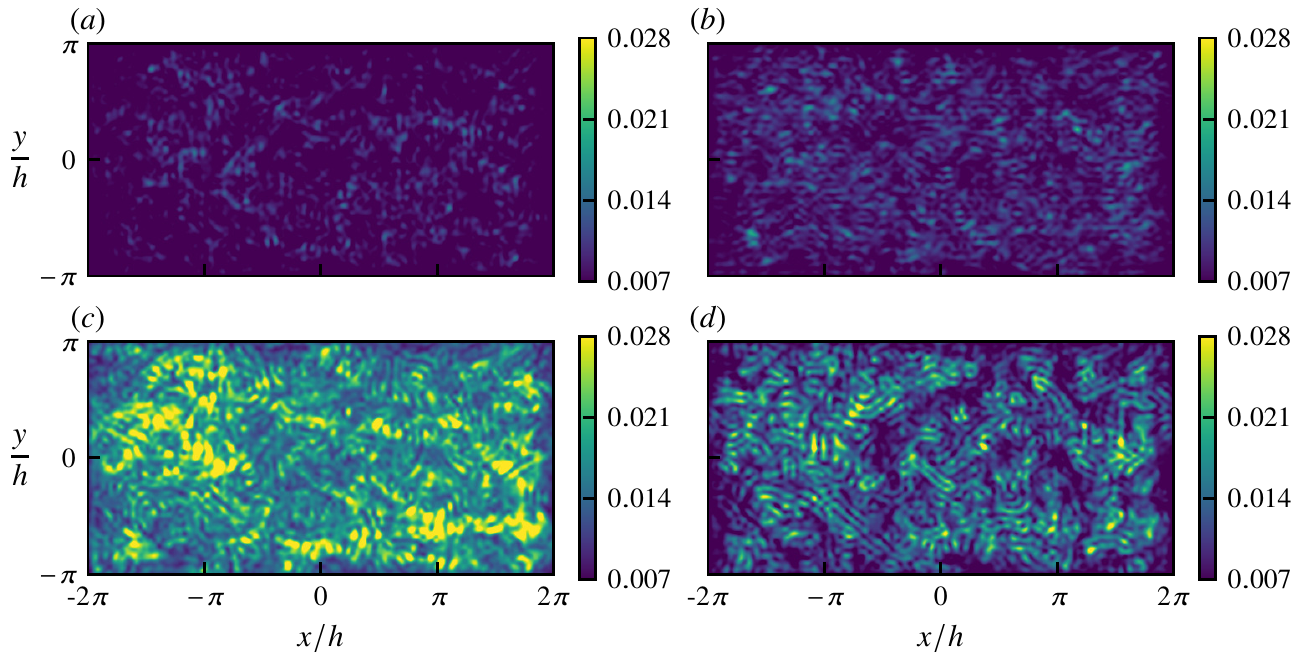}
    \caption{\label{fig:saliency_f008}Saliency maps for (\sublabel{a}) $u_s$, (\sublabel{b}) $v_s$, (\sublabel{c}) $w_s$ and (\sublabel{d}) $\eta$ at the surface of the flow with ${Fr}_\tau=0.08$.}
\end{figure}

\begin{figure}
    \centering
    \includegraphics{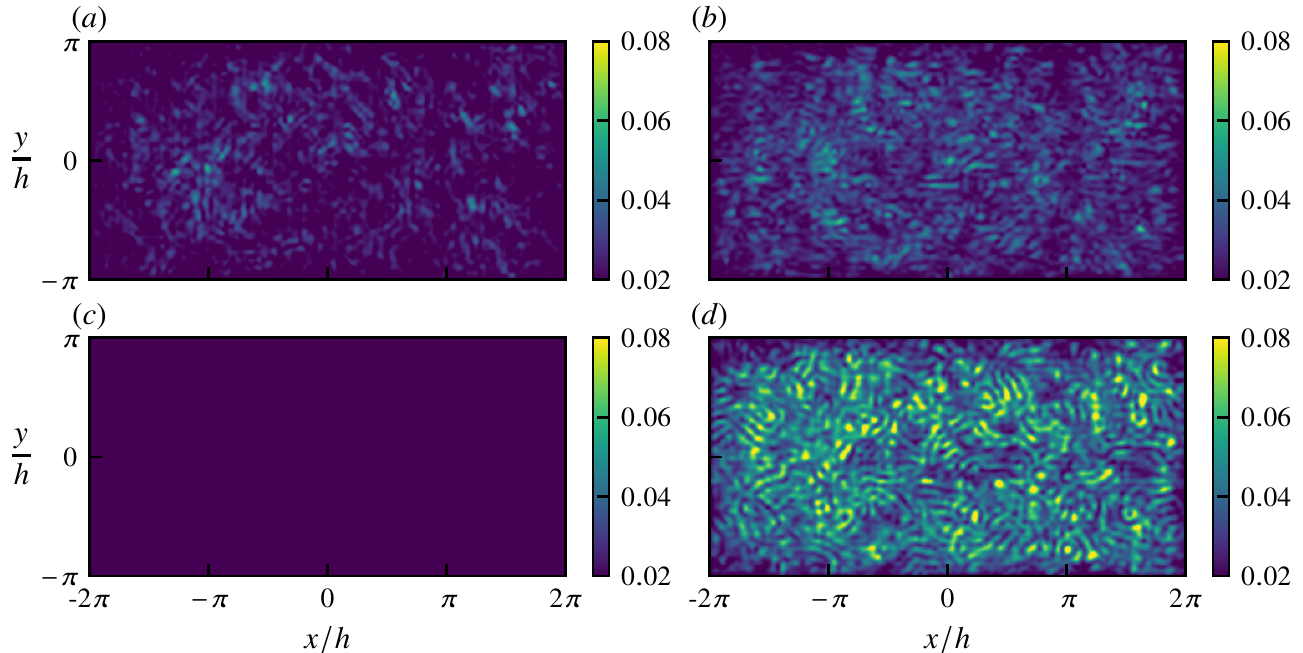}
    \caption{\label{fig:saliency_f001}Saliency maps for (\sublabel{a}) $u_s$, (\sublabel{b}) $v_s$, (\sublabel{c}) $w_s$ and (\sublabel{d}) $\eta$ at the surface of the flow with ${Fr}_\tau=0.01$.}
\end{figure}

Figures~\ref{fig:saliency_f008} and~\ref{fig:saliency_f001} show the saliency maps obtained for individual snapshots of the flows with $Fr_\tau=0.08$ and $Fr_\tau=0.01$, respectively.
For the flow with the high Froude number, the higher saliency values of the surface elevation $\eta$ and vertical velocity $w_s$ indicate that they are significant for the reconstruction process. By comparison, the saliency maps for the flow with the low Froude number indicate that the surface elevation $\eta$ and the spanwise velocity $v_s$ are more important than $u_s$ and $w_s$.

We can see that the surface elevation $\eta$ plays a more important role in the CNN method than in the LSE method, which indicates that the surface elevation contains useful information for inferring subsurface structures; however, the relations between $\eta$ and the subsurface flow field are likely complex and nonlinear, and, thus, cannot be captured by the LSE method.

\begin{figure}
    \centering
    \includegraphics{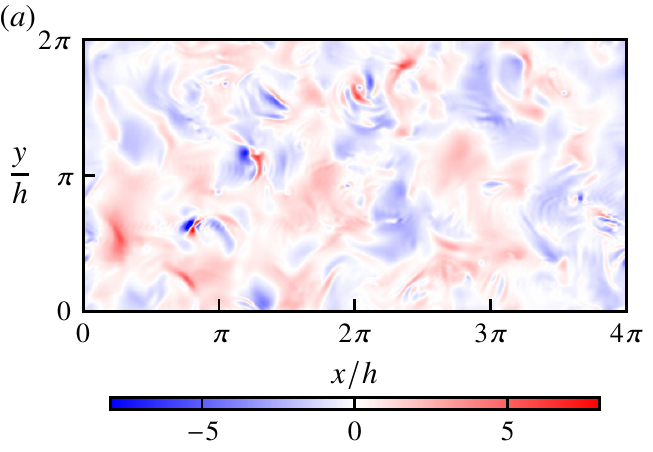}
    \includegraphics{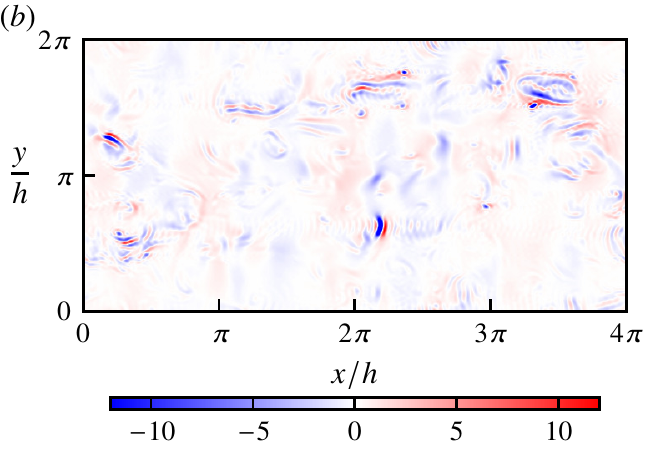}
    \caption{\label{fig:surface_w}Instantaneous vertical velocity fluctuations at the surface $w_s/w_{rms}$ for (\sublabel{a}) the flow with the high Froude number (${Fr}_\tau=0.08$) and (\sublabel{b}) the flow with the low Froude number (${Fr}_\tau=0.01$), where $w_{rms}$ is the root-mean-square value of $w_s$.}
\end{figure}

Another phenomenon specific to the CNN method is that the Froude number changes the importance levels of variables, which we do not observe in the kernels of the LSE method. Specifically, the vertical velocity fluctuations $w_s$ are affected: in the flow with the low Froude number, $w_s$ barely has any significance (figure~\ref{fig:saliency_f001}\sublabel{c}), which contrasts with its important role in the flow with the high Froude number (figure~\ref{fig:saliency_f008}\sublabel{c}).
This result suggests that the increased Froude number changes the behaviour of $w_s$, which means that different free-surface dynamics are present and need to be considered by the CNN when predicting subsurface flow structures. The vertical surface velocity fluctuations $w_s$ in the flows with low and high Froude numbers are plotted in figure~\ref{fig:surface_w}. In the flow with the high Froude number (figure~\ref{fig:surface_w}\sublabel{a}), the vertical surface motions are characterised by small-scale scars~\citep{sarpkaya1996,brocchini2001} and patches with larger scales. We note that these patchy structures are not obvious when moving slightly away from the surface, but scar-like structures are still present (see figure~\ref{fig:comparison_velocity_plane_w}\sublabel{a}), which indicates that the surface scars are directly induced by the subsurface flow structures, whereas the patchy motions are only significant near the surface and are likely governed by the free-surface dynamics. Figure~\ref{fig:surface_w}(\sublabel{b}) shows that in the flow with the low Froude number, the small-scale scars are dominant, indicating that the patchy motions only appear in the flow with the high Froude number. Therefore, we believe that the more deformable free surface in the flow with the high Froude number flow allows richer vertical velocity fluctuation structures near the surface, and as a result, the CNN needs the information of $w_s$ to reconstruct the near-surface flow structures. We conjecture that this change is related to the turbulence-induced roughness and the oscillatory motions induced by surface gravity waves~\citep{guo2010,savelsberg2008,dolcetti2016}. An analysis of the spatio--temporal spectrum of the free-surface fluctuations by~\citet{yoshimura2020} showed that at low Froude numbers, the free-surface fluctuations are mostly excited by subsurface turbulence structures, while at high Froude numbers, the free-surface motions consist of both turbulence-induced fluctuations and waves. The same trend is also observed in our dataset (not plotted). We note that the velocity structure is dependent on the nature of the free-surface fluctuations; therefore, it is expected that the characteristic features of the surface velocity fluctuations become different when waves are generated in the flow with the high Froude number.

\begin{figure}
    \centering
    \includegraphics{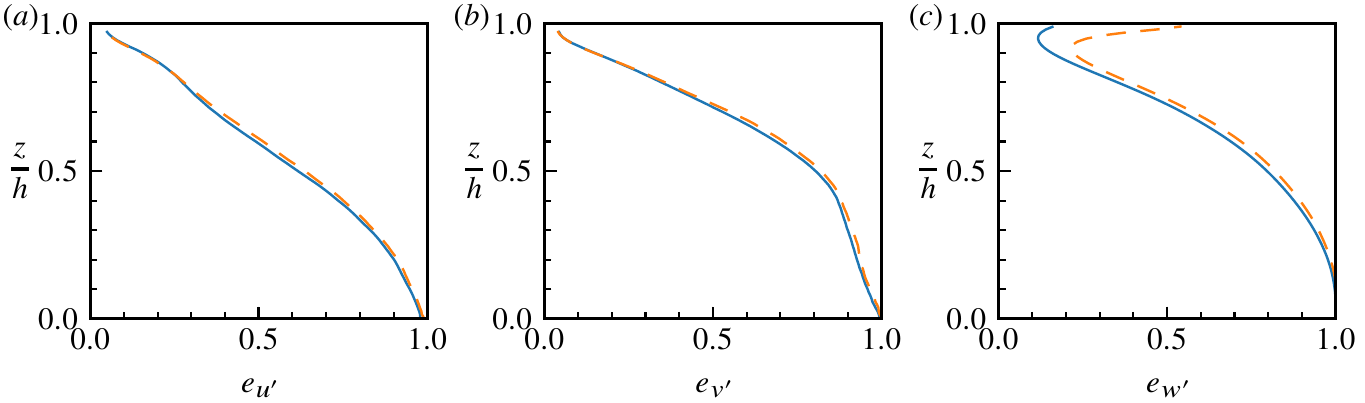}
    \includegraphics{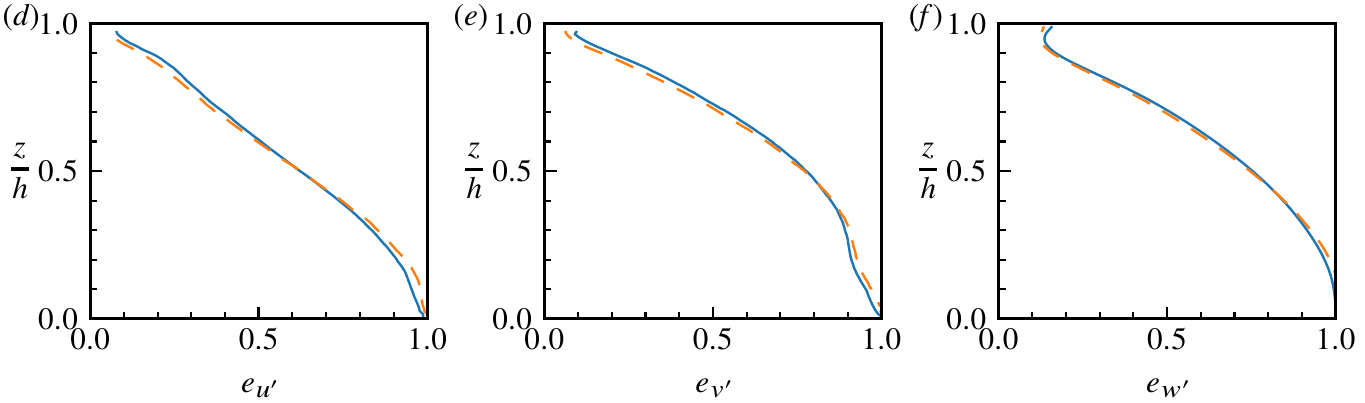}
    \caption{\label{fig:corr_coeff_without_w}\rev{Comparison between the normalised mean squared reconstruction errors of the CNN models with (\textcolor{colorC0}{\full}) and without (\textcolor{colorC1}{\broken}) $w_s$ included in the input. The upper (\sublabel{a}--\sublabel{c}) and lower (\sublabel{d}--\sublabel{f}) rows show the case of the high Froude number (${Fr}_\tau=0.08$) and the case of the low Froude number (${Fr}_\tau=0.01$), respectively, for the streamwise (\sublabel{a},\sublabel{d}), spanwise (\sublabel{b},\sublabel{e}) and vertical (\sublabel{c},\sublabel{f}) velocity fluctuations.}}
\end{figure}

To further understand the effects of $w_s$ on the flow reconstruction, we perform a numerical experiment by training a CNN without $w_s$ in the input. Figure~\ref{fig:corr_coeff_without_w} compares the performance of the CNN without $w_s$, \rev{measured by the normalised mean squared errors~\eqref{eq:l2_error_def}}, with that of the full model that uses $w_s$. We find that the reconstruction performance achieved for the flow with the low Froude number is barely affected by the missing $w_s$, which is consistent with our conclusion drawn from the saliency map that the reconstruction of the flow with the low Froude number barely depends on the information of $w_s$. For the flow with the high Froude number, removing $w_s$ from the input slightly degrades the reconstructions of the streamwise and spanwise velocities but has a notable impact on the reconstruction of the vertical velocity. \rev{Near the surface, the normalised reconstruction error increases to more than $0.5$ without $w_s$ compared to $0.18$ with $w_s$.} This result confirms our conclusion above that in flows with high Froude numbers, the vertical velocity fluctuations at the surface are associated with flow structures that are mostly concentrated near the surface.

\begin{figure}
    \centering
    \includegraphics{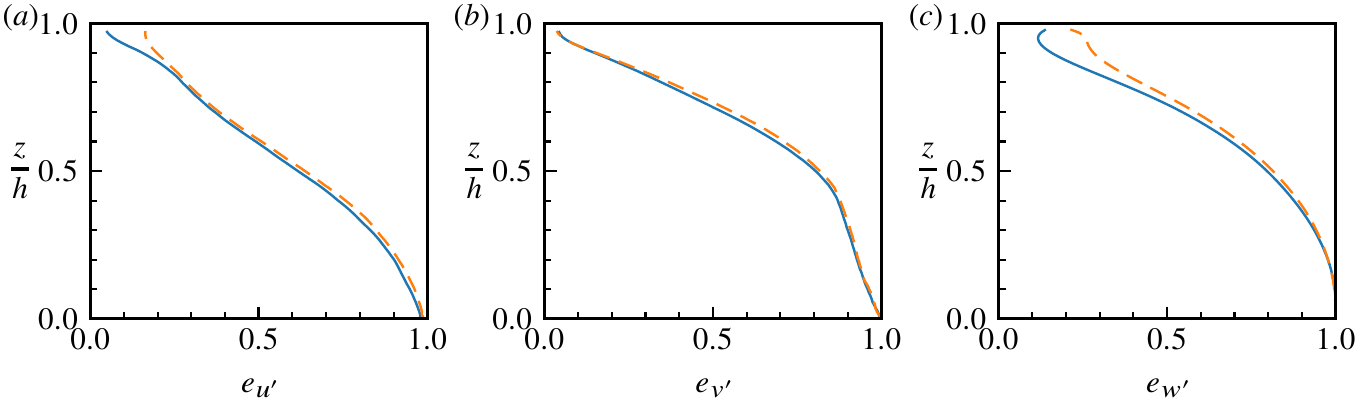}
    \caption{\label{fig:corr_coeff_without_u}\rev{Comparison between the normalised mean squared reconstruction errors of the CNN models with (\full) and without (\broken) $u_s$ included as input. The reconstruction errors of the (\sublabel{a}) streamwise, (\sublabel{b}) spanwise and (\sublabel{c}) vertical velocity fluctuations are plotted for the case with ${Fr}_\tau=0.08$.}}
\end{figure}

The saliency maps also indicate that the streamwise velocity at the surface $u_s$ is less important to flow reconstruction. Specifically, it is the least significant input for the case with ${Fr}_\tau=0.08$ (figure~\ref{fig:saliency_f008}) and the second-least significant input for the case with ${Fr}_\tau=0.01$ (figure~\ref{fig:saliency_f001}). To further confirm this conclusion, we train another network that removes $u_s$ from its inputs. \rev{The reconstruction error obtained for the case with ${Fr}_\tau=0.08$ is shown in figure~\ref{fig:corr_coeff_without_u}. Although the reconstruction produced without the information of $u_s$ is inferior, the difference is small and the overall predictions are still generally accurate.} These results are consistent with what the saliency maps show above.

The above analyses based on the saliency maps reveal that the information provided by the different surface variables is not equally important for the CNN when reconstructing subsurface flows; this finding is supported by the numerical experiments presented above. \rev{Moreover, the saliency maps can be considered as quantitative indicators of the surface--subsurface correlations. The fact that the CNN and LSE methods are sensitive to different surface variables suggests that the two methods give different interpretations about the physical relations between the free-surface motions and the subsurface velocities. Considering that the CNN model, which can describe nonlinear relations, has better reconstruction accuracy than the LSE model, we expect that the CNN model's description of the surface--subsurface relations is closer to the underlying physics. Indeed, some findings in the literature about the free-surface manifestations of the turbulence structures are not as clearly shown in LSE as in CNN.  For example, the surface elevation $\eta$ is known to be affected by the subsurface motions, which is supported by the spatial--temporal spectrum of $\eta$ showing a ridge corresponding to the advection of turbulence structures~\citep{savelsberg2008,dolcetti2016,yoshimura2020}. The relation of $\eta$ with the subsurface velocities is highlighted more by the CNN model than by the LSE model. For the LSE method, only limited surface--subsurface relations can be described by linear relations and, as a result, the model is sensitive to different surface variables. Specifically, according to the LSE model, $u_s$ and $v_s$ have strong linear correlations with near-surface motions. However, such correlations may not be as useful in the CNN model as in the LSE model because they are only valid near the surface and the nonlinear relations concerning other variables discovered by the CNN model may be more effective for the reconstruction.} Lastly, we note that the saliency maps indicate that the Froude number has a significant impact on the free-surface motions and, thus, can affect the information needed for reconstructions. The effect of the Froude number on the CNN performance is further investigated in~\S\,\ref{sec:generalization}.

The saliency maps also contain information about which surface regions contribute most to the reconstruction process. However, owing to the chaotic nature of turbulence, the salient regions become complex and difficult to interpret. \rev{First, we notice that the high salient regions feature a considerable amount of small-scale structures with localised points and grating-like shapes. We believe that the shapes of these salient regions indicate that the CNN samples surface features with certain spatial frequencies. The presence of the salient regions with relatively high spatial frequencies suggests that small-scale structures still have influences on flow reconstructions despite the fact that most reconstructed structures are large-scale motions. To further understand the influence of the small-scale surface features on the reconstructions, we reduce the surface input dimensions to $64 \times 32$ and observe increases in the reconstruction errors. Notably, the error in $w'$ near the surface is affected more than $u'$ and $v'$, which agrees with the observation that $w'$ features more small-scale structures than $u'$ and $v'$ in the near-surface region (see figures~\ref{fig:comparison_velocity_plane_w}\sublabel{a}, \ref{fig:comparison_velocity_plane_u}\sublabel{a}, and~\ref{fig:comparison_velocity_plane_v}\sublabel{a}). The reduction in the reconstruction accuracy is less obvious away from the surface, indicating that small-scale surface features mostly affect the reconstructions of small-scale vertical fluctuations near the surface. It should also be noted that the distribution of the salient regions still has the signatures of large-scale structures. The following is an example that we discover supporting the existence of correlations between the salient regions and the known flow features underneath the free surface.}

\begin{figure}
    \centering
    \includegraphics{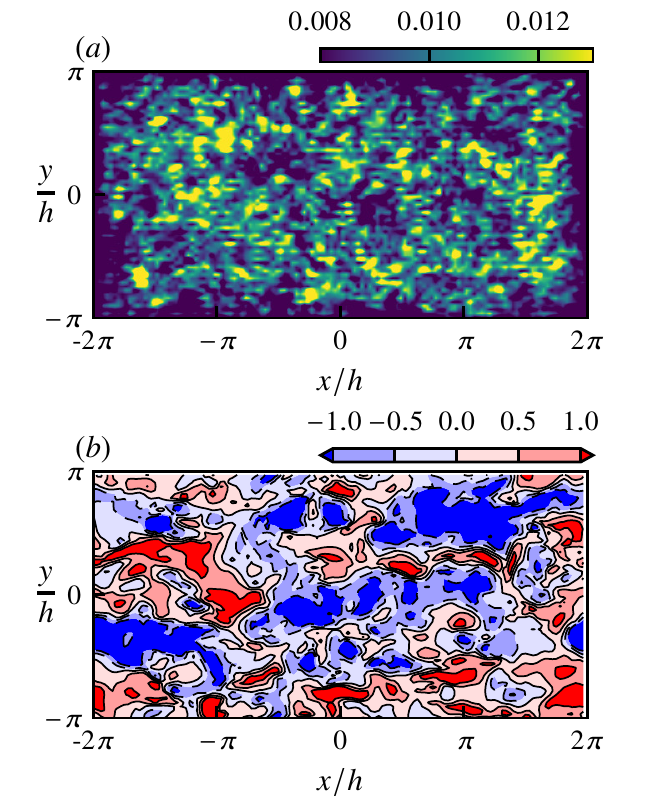}
    \includegraphics{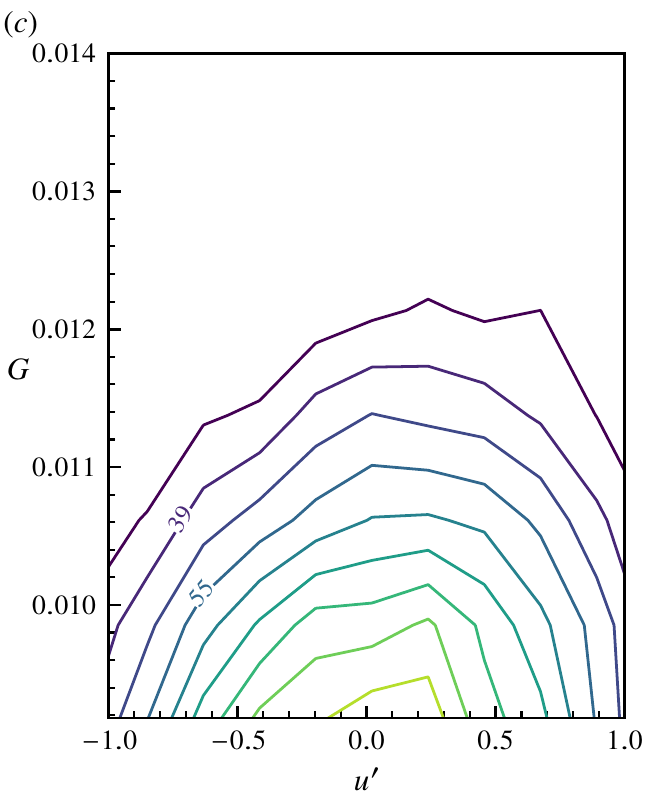}
    \caption{\label{fig:saliency_correlation_v}Relations between (\sublabel{a}) the salient regions of $v_s$ and (\sublabel{b}) the subsurface $u'$ at $z/h=0.9$. The dashed contours in (\sublabel{b}) indicate negative $u'$ values. In (\sublabel{c}), the contours of the JPDF of the saliency $G$ and $u'$ are plotted. To highlight the relationship between the salient regions with high $G$ and $u'$ values, the contours are cut off at the median of $G$; i.e.\ only the upper half of $G$ is plotted.}
\end{figure}

Figures~\ref{fig:saliency_correlation_v}(\sublabel{a}) and~\ref{fig:saliency_correlation_v}(\sublabel{b}) show the saliency map of $v_s$ and the contours of $u'$ on a subsurface plane at $z/h=0.9$, respectively, and we find that a correlation is present between the salient regions of $v_s$ and the regions of positive $u'$ values. To examine this relationship, we plot the joint probability density function (JPDF) of the saliency $G$ and $u'$ in figure~\ref{fig:saliency_correlation_v}(\sublabel{c}). The direction of the JPDF contours indicates that a higher saliency value is more likely to coincide with positive $u'$ values. We note that the streaky structure of the streamwise fluctuations is one of the characteristics of turbulent shear flows and that the CNN can capture large-scale streaks away from the surface. Therefore, we conjecture that the coincidence of the salient regions of $v_s$ with positive $u'$ values below the surface contributes to the CNN's ability to reconstruct the streamwise streaks.

The correlations between the salient regions and the known characteristic structures indicate that even though the turbulent structures are complex, they have characteristic free-surface manifestations that make them identifiable or reconstructable. In other words, the CNN method reconstructs the flow of interest by identifying critical surface regions that correspond to dynamically important structures under water. We note that critical regions do not necessarily correspond to large values of $\eta$ or $\boldsymbol{u}_s$; this behaviour is different from the LSE model, which assumes that the subsurface structures are linearly dependent on the surface quantities.

The above analyses indicate that although NNs are sometimes considered `black boxes', it is possible to obtain information about the flow physics of free-surface turbulent flows from such networks. In the present study, saliency maps, which we consider a starting point for understanding how the CNN utilises surface quantities, can indicate how the Froude number changes the relations between the free-surface motions and the subsurface flow structures and where strong manifestations of the subsurface flow structures exist. This information may be useful for studying the interaction between turbulence and a free surface.

\subsection{\label{sec:generalization}Generalization of CNNs to flows with different Froude numbers}
Each of the CNN models in the preceding sections is trained and tested with the same Froude number. In this section, the generalization abilities of the CNNs for addressing different Froude numbers are investigated. We focus on what performance can be attained when a model trained at a different Froude number is applied to flow reconstruction. Two scenarios are considered: predicting flows with lower Froude numbers using the model trained at a high Froude number (${Fr}_\tau=0.08$) and predicting flows with higher Froude number flows using the model trained at a low Froude number (${Fr}_\tau=0.01$).

\begin{figure}
    \centering
    \includegraphics[width=0.99\textwidth]{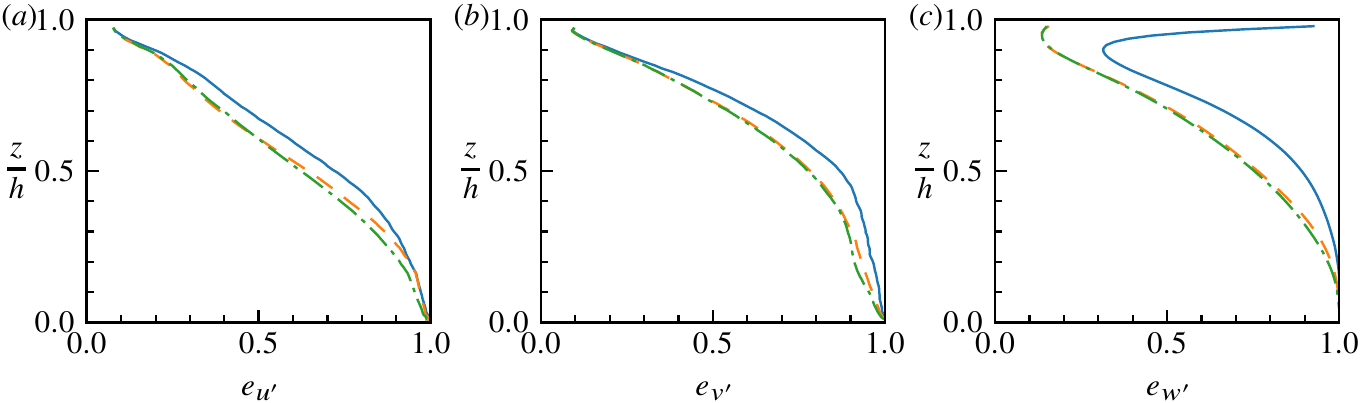}
    \includegraphics[width=0.99\textwidth]{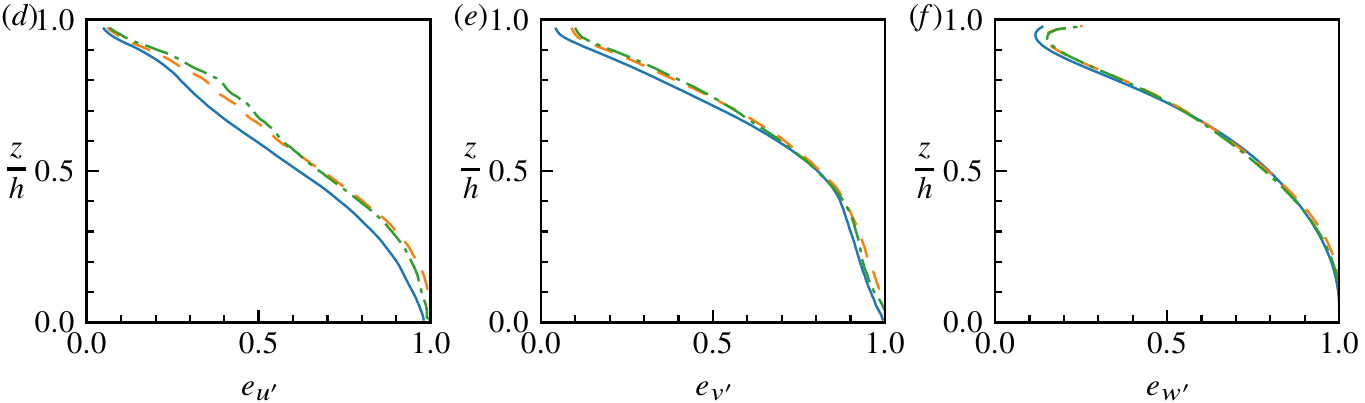}
    \caption{\label{fig:l2_cross_model}\rev{Normalised mean squared reconstruction errors when (\sublabel{a}--\sublabel{c}) the CNN trained at ${Fr}_\tau=0.01$ and (\sublabel{d}--\sublabel{f}) the CNN trained at ${Fr}_\tau=0.08$ are used to predict flows with different ${Fr}_\tau$. The Froude numbers of the flows to be reconstructed include ${Fr}_\tau=0.08$ (\textcolor{colorC0}{\full}), ${Fr}_\tau=0.03$ (\textcolor{colorC1}{\broken}) and ${Fr}_\tau=0.01$ (\textcolor{colorC2}{\chain}). The errors of the (\sublabel{a}) streamwise, (\sublabel{b}) spanwise and (\sublabel{c}) vertical velocity fluctuations are plotted.}}
\end{figure}

The reconstruction performance achieved in the two scenarios is shown in figure~\ref{fig:l2_cross_model}. We find that for the CNN model trained at ${Fr}_\tau=0.01$ (figure~\ref{fig:l2_cross_model}\sublabel{a}--\sublabel{c}), its reconstruction of the flow at ${Fr}_\tau=0.03$ is almost as accurate as its reconstruction of the flow at the Froude number used for training. However, when the network is applied to predict the flow at an even higher Froude number, ${Fr}_\tau=0.08$, its performance decreases considerably. By comparison, when we apply the CNN model trained at ${Fr}_\tau=0.08$ to flows with lower Froude number flows, we observe more consistent performance across different cases (figure~\ref{fig:l2_cross_model}\sublabel{d}--\sublabel{f}). The accuracy of the reconstructed $u'$ at ${Fr}_\tau=0.03$ and ${Fr}_\tau=0.01$ is slightly worse than the accuracy at the trained ${Fr}_\tau=0.08$, whereas the accuracy of $v'$ and $w'$ barely changes under various ${Fr}_\tau$.

\begin{figure}
    \centering
    \includegraphics[width=0.99\textwidth]{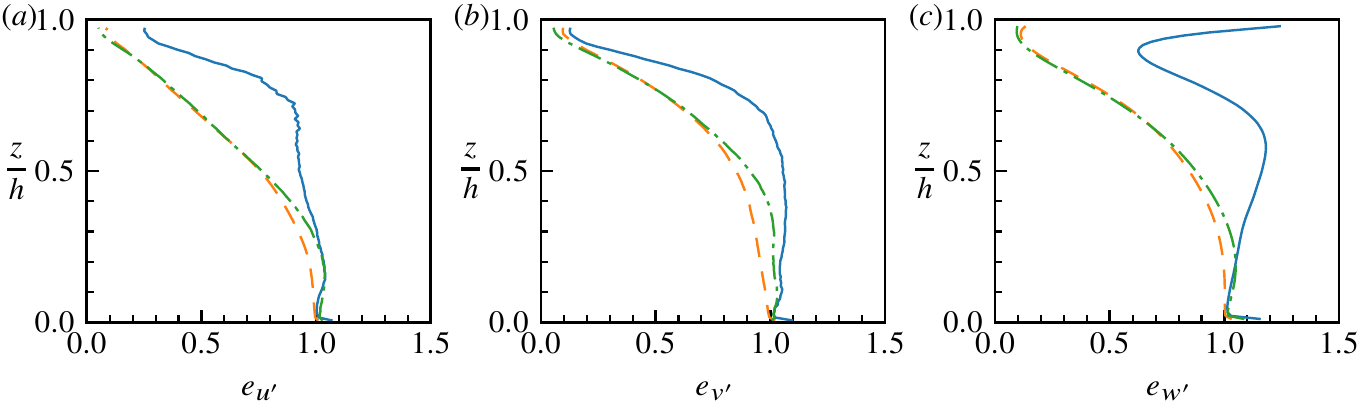}
    \caption{\label{fig:l2_cross_model_without_eta}\rev{Normalised mean squared reconstruction errors when a CNN model trained at ${Fr}_\tau=0.01$ without $\eta$ included as input is used to predict flows with different ${Fr}_\tau$. The Froude numbers of the flows to be reconstructed include ${Fr}_\tau=0.08$ (\textcolor{colorC0}{\full}), ${Fr}_\tau=0.03$ (\textcolor{colorC1}{\broken}) and ${Fr}_\tau=0.01$ (\textcolor{colorC2}{\chain}). The errors of the (\sublabel{a}) streamwise, (\sublabel{b}) spanwise and (\sublabel{c}) vertical velocity fluctuations are plotted.}}
\end{figure}

The above results indicate that the CNN model trained at a low Froude number cannot fully describe the structures present in flows with high Froude numbers. We note that the preceding analyses concerning the saliency of the CNN suggest that one of the distinct differences between the reconstructions of flows with low Froude numbers and those with high Froude numbers is that the vertical surface fluctuations are unimportant when predicting the former type but are important for the latter type, especially for the vertical fluctuations near the surface. Figure~\ref{fig:l2_cross_model}(\sublabel{c}) shows that the accuracy of the predicted $w'$ declines sharply near the surface, indicating that the poor performance of the model is related to its inability to process the structures of $w_s$.

\rev{To further confirm that the poor generalization capability of the low Froude number model is related to the changes in the $w_s$ structures, we train a low Froude number model without $\eta$ in the input to force the model to rely only on surface velocities to infer subsurface flows. This test is inspired by the observation that $\eta$ and $w_s$ in the LSE seem to provide similar information (figure~\ref{fig:lse_kernel_f08_09}) but $w_s$ is neglected by the low Froude number model. We hope that removing the information of $\eta$ may force the CNN model to use $w_s$ for reconstructions. The reconstruction errors are plotted in figure~\ref{fig:l2_cross_model_without_eta} and exhibit increases for all cases, indicating that the information of the surface elevation is useful for CNN reconstructions, consistent with the analyses based on saliency maps (figure~\ref{fig:saliency_f001}). Comparing the performances for different cases, we notice that the reconstruction errors increase significantly at ${Fr}_\tau=0.08$ and the largest errors are still present in the vertical fluctuations, similar to what we observe in figure~\ref{fig:l2_cross_model}(\sublabel{a}--\sublabel{c}). Near the surface, the normalised error $e_{w'}$ is even greater than $1$. This result further confirms that the structures of $w_s$ at ${Fr}_\tau=0.08$ unseen by the low Froude number model contribute to its poor generalization performance.}

Interestingly, the model trained at a high Froude number is quite versatile and can be applied to flows with lower Froude numbers while achieving only slightly degraded performance. This result indicates that even though the CNN is optimised for the processes in flows with high Froude numbers, it still maintains the ability to predict the common relations in the flows with both high and low Froude numbers and reconstruct the flow field consistently without tuning.

\rev{At last, considering the normalisation of the input variables as introduced in~\S\,\ref{sec:preprocess}, we can gain further insights into the role of the surface elevation $\eta$ in the CNN model. The surface elevation $\eta$ is normalised by its root-mean-square value when fed into the CNN. In other words, only the geometry structures of $\eta$ are input into the CNN despite that the unnormalised $\eta$ varies by several orders of magnitude from $Fr_\tau=0.01$ to $Fr_\tau=0.08$ (see figure~\ref{fig:comparison_etarms}). The good generalization capability of the CNN model indicates that only the geometry feature, rather than the magnitudes of the surface elevations, are important to the subsurface flow reconstructions.}

\rev{
\subsection{\label{sec:size_generalization}Generalization of CNNs to variable sized inputs}
Lastly, we discuss the applications of the CNN model to variable sized inputs. Specifically, two aspects are considered, the spatial resolution of inputs and the domain size.

Because the training is performed with inputs with a fixed resolution, the trained network can only process inputs at this resolution.
When the input resolution is lower than the resolution that the model is trained at, one can upsample the input to the required resolution through interpolations as what is commonly done in image processing. However, the features added by the interpolations may not be physical and can negatively impact the reconstructions. In the test reported in \S\,3 of the supplementary material, we find that the performance of using interpolated inputs with our trained model is inferior to the performance of a model trained with the low resolution input. Therefore, interpolations should be used with care. This issue can be resolved by training the CNN with a variety of resolutions to improve the model's generalization capability but this is beyond the scope of this paper. One can also adapt the encoder to the desired input resolution, reuse the decoder architecture and its trained weights, and use transfer learning to obtain a new model with reduced training data and time~\citep{pan2010}.

For domain sizes, because the proposed CNN model only uses convolution, downsampling, and upsampling operations, the network can process inputs with different dimensions. The resulting intermediate feature maps and the final output are scaled proportionally. This type of CNN can be categorized as fully convolutional neural networks, which are widely used for processing variable sized images~\citep*{long2015}. This design provides some generalization capability concerning variable sized inputs. Moreover, because the simulation domain used in the present study is large enough to contain most scales that are present in open-channel flows (see discussions in~\S\,\ref{sec:dataset}), the flow structures in a larger domain should not differ much from the learned features in the present model and, therefore, the kernel weights can be reused for inputs with larger domain sizes. However, if the input domain size is smaller than the present domain size, e.g. only measurements in a partial domain are available, the flow structures and their corresponding surface manifestations may be trimmed at the boundaries. The present CNN model is not developed for handling this scenario and a new model is needed and transfer learning can also be used to develop the new model more efficiently.}

\section{\label{sec:conclusions}Conclusions}
This study presents a CNN designed to reconstruct the subsurface flow velocity of a turbulent open-channel flow using the surface information and investigates the relationship of the reconstruction performance to the flow physics. The CNN model consists of an encoder that extracts information from a two-dimensional discrete grid of surface variables, including the surface elevation and the fluctuating surface velocities, and a decoder that reconstructs a three-dimensional velocity field from the extracted surface information. We find that the CNN method can infer the near-surface velocities with high accuracy, as verified by both a visual inspection of instantaneous fields and \rev{quantitative measures using normalised mean squared errors. The amplitude spectra and the phase errors of different Fourier modes are also analysed to assess the scale-specific reconstruction performance.} Away from the surface, the accuracies of the reconstructions decrease, but the CNN method retains good accuracies when predicting low-wavenumber structures, specifically the large-scale streamwise streaks in the open-channel flow, even down to the lower half of the channel. We also consider a conventional reconstruction method based on LSE, which produces significantly worse reconstructions away from the surface than the CNN method and provides good reconstructions only near the surface. The performance assessments indicate that the proposed CNN model is an effective tool for inferring the subsurface flow field and by using this method, a considerable amount of subsurface flow information, including the three-dimensional velocities of certain large-scale flow structures in the lower half of the channel, can be inferred from free surfaces.

In addition to the thorough evaluation of the performance expectations regarding the subsurface flow inference, we further analyse the LSE and CNN methods to understand how both methods reconstruct the subsurface flow field. For the LSE model, the kernels that map the surface quantities to the subsurface velocities reveal that there exist linear correlations between the horizontal surface velocities and the subsurface vortices, which the LSE method uses to reconstruct the subsurface flow field. For the CNN model, we use saliency maps to investigate what surface information is more important for reconstructing subsurface flows. The correlations between the important surface regions identified by the saliency values and the characteristic subsurface flow features indicate that the CNN relies on identifying the surface signatures of subsurface flow structures to reconstruct the flow field. We also find that some surface variables play lesser roles in the reconstruction process, but the specific variables are affected by the Froude number and the associated change of free-surface dynamics. Specifically, when addressing flows with low Froude numbers, the vertical surface velocity fluctuations can almost be neglected, but they become important when reconstructing flows with high Froude numbers, owing to the presence of different near-surface vertical velocity structures. In addition, the generalization capability of the CNN model with regard to the Froude number is assessed. It is found that the CNN model trained at a high Froude number can be applied to flows with lower Froude numbers and achieve good accuracy. However, the CNN model trained at a low Froude number has a limited capability of reconstructing flows with high Froude numbers owing to missing physical relations such as the motions associated with the vertical fluctuations of the free surface. These analyses provide physical insights into the mechanisms underlying the reconstruction process and the effect of free-surface flow dynamics on the reconstruction process.

The present study affirms that the CNN is a promising technique for the inverse problems concerning free-surface flows. This is not just because the CNN achieves a good performance for subsurface flow inference applications; more importantly, this work demonstrates that physical insights can be obtained from the CNN model and, therefore, the CNN can assist with the understanding of free-surface flow dynamics.
Specifically, we consider that the CNN can reveal physical processes underlying the interactions between turbulent flows with free surfaces from two aspects. First, the reconstructed flow field contains the subsurface flow structures that can influence free-surface motions, and the ability of the CNN to describe nonlinear mappings can enable it to discover flow structures that affect the free surface through nonlinear processes. For example, in the present study, based on the near-wall streaks present in the reconstructed flow field (figure~\ref{fig:comparison_velocity_plane_u}\sublabel{h}), we can determine that these streaks are related to free-surface motions; and the vortical structures derived from the reconstruction (figures~\ref{fig:vortex} and~\ref{fig:vortex_xz}) further reveal that the streaks and the free surface are connected through the inclined vortices. We note that such relations were traditionally discovered by directly observing of the evolution of the instantaneous flow field and conditional averaging~\citep[see, e.g.][]{rashidi1997,shen1999,nagaosa2003,sanjou2011}. \rev{By representing the surface--subsurface relations as a neural network, the reconstruction by the CNN provides a new approach of extracting characteristic structures that correlate with the free-surface motions. Different from existing methods such as the conditional averaging, the structures obtained by the CNN are instantaneous and can be tracked in time. Analysing these time-resolved reconstructed structures may provide further insights into the evolution dynamics of these structures and how they interact with the free surface, similar to the studies performed for turbulence channel flows~\citep[see e.g.][]{lozano-duran2014}.} Second, by analysing how the CNN model processes the inputs, one can gain further insights into the dynamics of free-surface turbulence. For example, the analysis conducted based on saliency maps, which shows that importance level of the inputs changes with the Froude number of the flow, leads us to find the change of the structures of vertical surface fluctuations with the Froude number. The surface salient regions reveal the important surface features identified by the CNN, which provide the possibility of studying the free-surface manifestations of flow structures. We believe that more opportunities for studying physical relations using CNNs are expected to emerge as more network interpretation methods are introduced~\citep{zhang2018b}.

\backsection[Declaration of interests]{The authors report no conflict of interest.}

\begin{figure}
    \centering
    \includegraphics{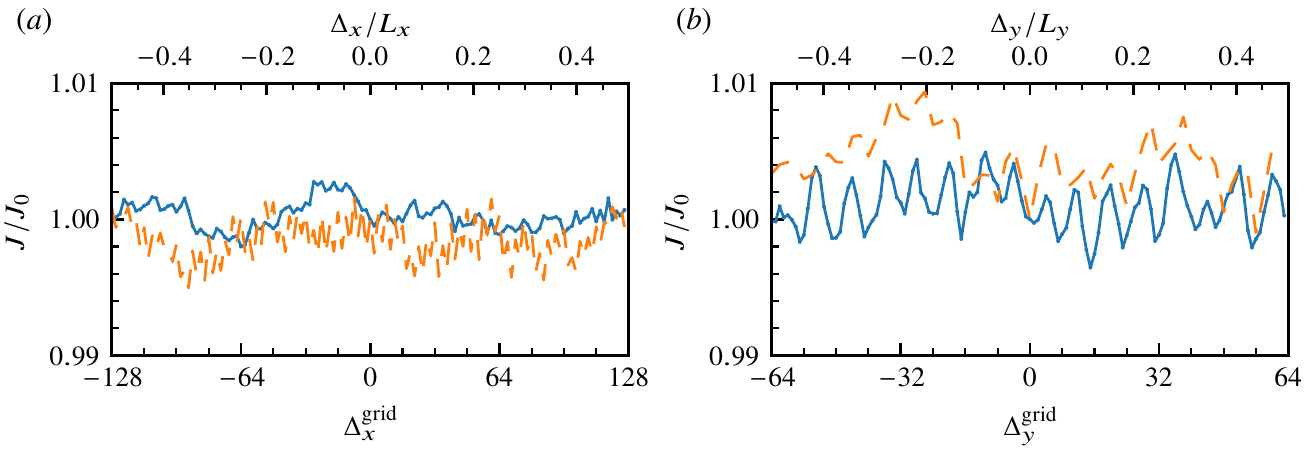}
    \caption{\label{fig:translation_error}\rev{Variations of the reconstruction loss $J$~\eqref{eq:loss} with the translation distances. The losses are normalised by the loss for a non-translated input, $J_0$. Two CNN models are considered: the original model as presented in table~\ref{tab:nn_param} with the blur pooling layers and periodic paddings(\textcolor{colorC0}{\full}) and a model with only strided convolution layers and zero paddings (\textcolor{colorC1}{\broken}). The translation distances in the (\sublabel{a}) $x$ and (\sublabel{b}) $y$ directions are denoted by $\Delta_x^\text{grid}$ and $\Delta_y^\text{grid}$ in terms of the number of grid points, and by $\Delta_x/L_x$ and $\Delta_y/L_y$ as relative distances to domain sizes, respectively.}}
\end{figure}

\appendix
\rev{\section{\label{sec:appendix_translation}Effects of translation on reconstruction accuracy}
To evaluate the CNN model's ability to maintain translation invariance, we compute the reconstructions from the periodically shifted inputs with different translation distances and compare how the reconstruction accuracy varies with the translation. The variations of the reconstruction loss $J$~\eqref{eq:loss} with the translation distances are plotted in figure~\ref{fig:translation_error}. We find that the variation in the reconstruction performance is less than $0.5\%$, indicating that the CNN model can produce equivalent reconstructions when the inputs are shifted and therefore should maintain a good translation invariance property when establishing the surface--subsurface mappings for reconstructions. For comparison, we also consider a CNN model that only uses strided convolutions for downsampling instead of the blur layer and uses the zero padding instead of the periodic padding. This model has a larger performance variation when the inputs are shifted. The above result indicates that the blur pooling layers and periodic paddings adopted in the present CNN model can indeed improve the model's translation invariance.}

\section{\label{sec:instantaneous_vw}Instantaneous spanwise and vertical velocity fluctuations}
\begin{figure}
    \includegraphics[width=0.99\textwidth]{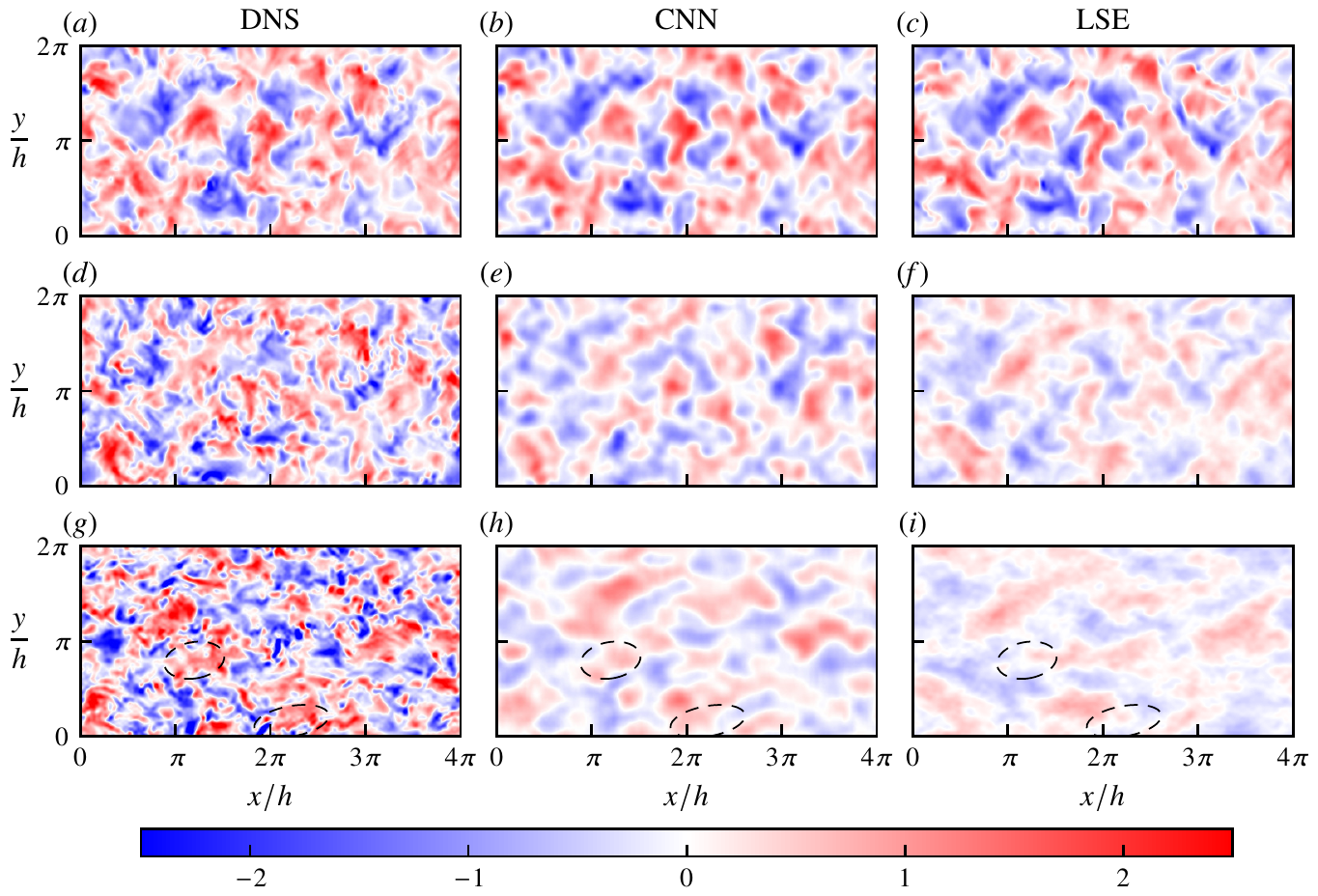}
    \caption{\label{fig:comparison_velocity_plane_v}Comparisons among the instantaneous spanwise velocity fluctuations $v'$ obtained from (\sublabel{a},\sublabel{d},\sublabel{g}) the DNS results and the fields reconstructed by (\sublabel{b},\sublabel{e},\sublabel{h}) the CNN and (\sublabel{c},\sublabel{f},\sublabel{i}) LSE methods. The $x$-$y$ planes at (\sublabel{a}--\sublabel{c}) $z/h=0.9$, (\sublabel{d}--\sublabel{f}) $z/h=0.6$ and (\sublabel{g}--\sublabel{i}) $z/h=0.3$ are plotted for the case of ${Fr}_\tau=0.08$.}
\end{figure}

\begin{figure}
    \includegraphics[width=0.99\textwidth]{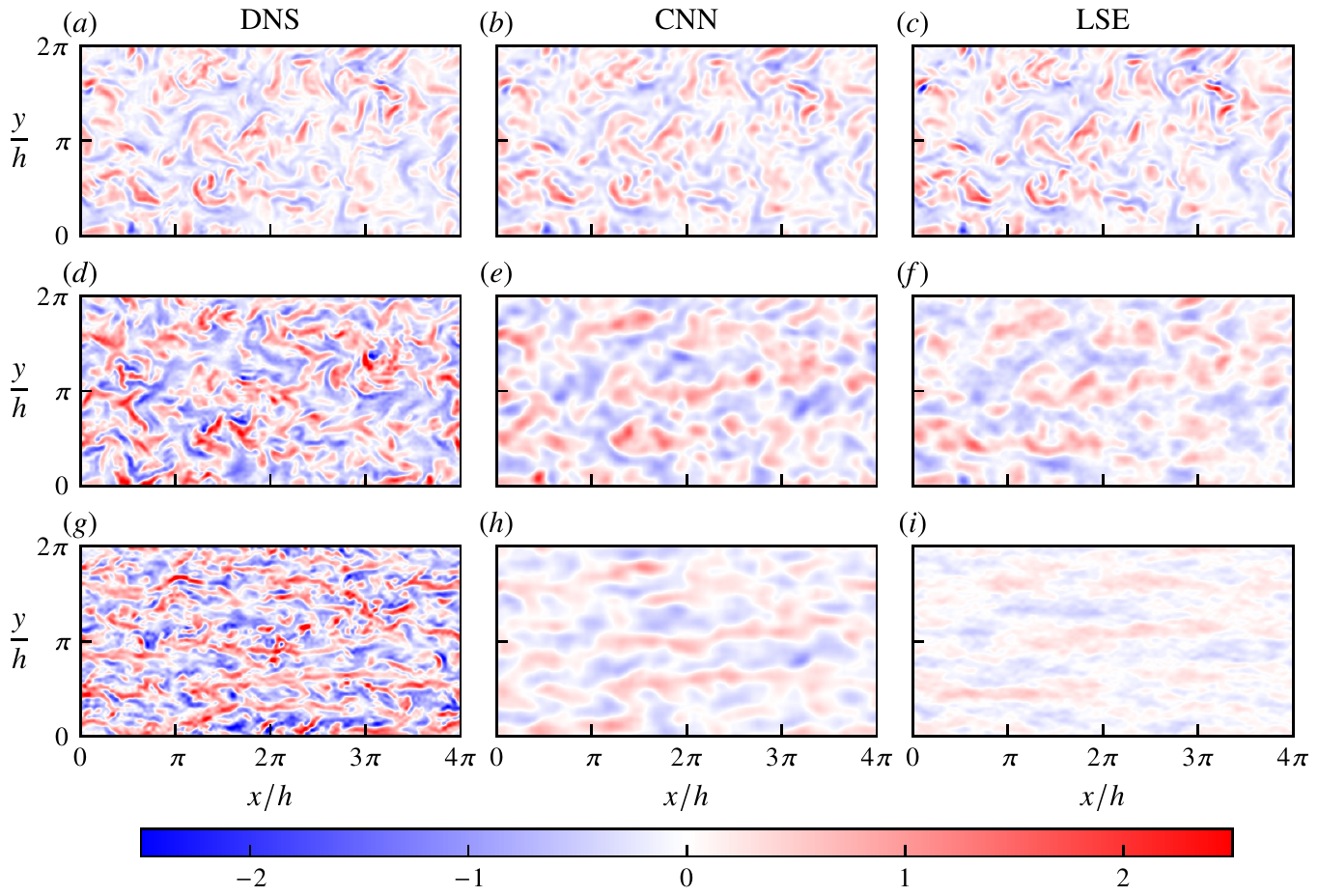}
    \caption{\label{fig:comparison_velocity_plane_w}Comparisons among the instantaneous vertical velocity fluctuations $w'$ obtained from (\sublabel{a},\sublabel{d},\sublabel{g}) the DNS results and the fields reconstructed by (\sublabel{b},\sublabel{e},\sublabel{h}) the CNN and (\sublabel{c},\sublabel{f},\sublabel{i}) LSE methods. The $x$-$y$ planes at (\sublabel{a}--\sublabel{c}) $z/h=0.9$, (\sublabel{d}--\sublabel{f}) $z/h=0.6$ and (\sublabel{g}--\sublabel{i}) $z/h=0.3$ are plotted for the case of ${Fr}_\tau=0.08$.}
\end{figure}

The instantaneous spanwise and vertical velocity fluctuations reconstructed by the CNN and LSE methods are compared to those obtained from the DNS data in figures~\ref{fig:comparison_velocity_plane_v} and~\ref{fig:comparison_velocity_plane_w}, respectively. Overall, we observe that the CNN method can reconstruct the large-scale features of the spanwise and vertical velocities away from the surface more accurately than the LSE method; this is similar to our findings for the streamwise velocity fluctuations (figure~\ref{fig:comparison_velocity_plane_u}). In general, compared to the streamwise velocity fluctuations $u'$, the spanwise velocity fluctuations $v'$ have smaller streamwise scales in the near-wall regions, posing difficulties for both the CNN and the LSE methods. However, the CNN method can still capture features that are not present in the LSE method (see, e.g. the positive $v'$ regions marked by the dashed circles in figure~\ref{fig:comparison_velocity_plane_v}\sublabel{g}--\sublabel{i}). The vertical velocity fluctuations $w'$ also feature many fine-scale structures, which both the CNN and LSE methods can predict with good accuracy near the surface. Moving away from the surface, the fine-scale structures in the vertical velocities are missing from the reconstructed flow field. However, we note that near the wall, the vertical velocity fluctuations estimated by the CNN method exhibit the patterns of the streamwise streaks and are roughly negatively correlated with the streamwise velocities (figure~\ref{fig:comparison_velocity_plane_u}\sublabel{h}), corresponding to positive Reynolds shear stress values $-u'w'$. This result indicates that the CNN method captures the correlations between the streamwise streaks and the Reynolds shear stress, which is an expected feature of the turbulent shear flow above the wall. Overall, with consistent results for $u'$ and $v'$, the CNN method provides better reconstruction performance for $w'$ than the LSE method.

\bibliographystyle{jfm}
\bibliography{flow_reconstruction}

\end{document}